\documentclass[12pt]{iopart}

\usepackage{wrapfig}
\usepackage{graphicx}
\usepackage{amsmath}
\usepackage{xcolor}
\usepackage{comment}
\usepackage{amssymb}

\definecolor{C0}{RGB}{31,119,180}
\definecolor{C1}{RGB}{255,127,14}
\definecolor{C2}{RGB}{44,160,44}
\definecolor{C3}{RGB}{214,39,40}
\definecolor{C4}{RGB}{148,103,189}

\begin{document}

\title[database]{Physics insights from a large-scale 2D UEDGE simulation database for detachment control in KSTAR}

\author{M. Zhao$^1$, X. Xu$^1$, B. Zhu$^1$, T.D. Rognlien$^1$, X. Ma$^2$, W.H. Meyer$^1$, K. Kwon$^2$, D. Eldon$^2$, N. Li$^1$, H. Lee$^3$, J. Hwang$^4$}

\address{$^1$Lawrence Livermore National Laboratory, 7000 East Ave, Livermore, CA 94550, US}
\address{$^2$General Atomics, P.O. Box 85608 San Diego, CA 92186, US}
\address{$^3$Korea Institute of Fusion Energy, 169-148 Gwahak-ro, Yuseong-gu, Daejeon, 34133, Republic of Korea}
\address{$^4$Korea Advanced Institute of Science and Technology, 291 Daehak-ro, Yuseong-gu, Daejeon 34141, Republic of Korea}

\ead{zhao17@llnl.gov}
\vspace{10pt}
\begin{indented}
\item[]September 2025
\end{indented}

\begin{abstract}
A large-scale database of two-dimensional UEDGE simulations has been developed to study detachment physics in KSTAR and to support surrogate models for control applications. Nearly 70,000 steady-state solutions were generated, systematically scanning upstream density, input power, plasma current, impurity fraction, and anomalous transport coefficients, with magnetic and electric drifts across the magnetic field included. The database identifies robust detachment indicators, with strike-point electron temperature at detachment onset consistently $T_{e,\mathrm{target}} \sim 3{-}4$~eV, largely insensitive to upstream conditions. Scaling relations reveal weaker impurity sensitivity than one-dimensional models and show that heat flux widths follow Eich’s scaling only for uniform, low $D$ and $\chi$. Distinctive in--out divertor asymmetries are observed in KSTAR, differing qualitatively from DIII-D. 
Complementary time-dependent simulations quantify plasma response to gas puffing, with delays of $5–15$ ms at the outer strike point and $\sim 40$ ms for the low-magnetic-field-side (LFS) radiation front. These dynamics are well captured by first-order-plus-dead-time (FOPDT) models and are consistent with experimentally observed detachment-control behavior in KSTAR [Gupta et al., submitted to Plasma Phys. Control. Fusion (2025)]

\end{abstract}

%
%
%
%
%

\section{Introduction}

One of the critical challenges in the development of magnetic confinement fusion energy is the effective management of plasma heat exhaust at the edge of tokamak devices. The scrape-off layer (SOL), where the heat exhaust is handled, plays a crucial role in determining the performance and longevity of plasma-facing components (PFCs). In this region, plasma heat and particles are transported across the magnetic separatrix and directed toward the divertor target plates along magnetic field lines. The heat load deposited on the divertor plates should be carefully controlled to remain below the engineering limits \cite{10MW}.

Achieving and sustaining divertor plasma detachment \cite{detachment1,detachment2} is a key method to address this issue. In such detached state, the plasma near the divertor plate become sufficiently cold and dense that most of the energy and momentum are dissipated well before reaching the material surface. The heat flux to PFCs is drastically reduced through radiation cooling in this regime. Therefore, maintaining divertor detachment is essential for protecting PFCs, extending device lifetimes, and ensuring that the power exhaust remains within tolerable limits in tokamak fusion devices, especially in future high-power devices such as ITER and DEMO. However, maintaining stable detachment is nontrivial and requires active control of boundary plasma actuators, including impurity seeding and gas puffing. The actuator must be dynamically adjusted in real time to maintain an optimal level of radiation in the divertor region. If the radiated power is insufficient, the plasma can reattach to the plates, exposing the target plates to high heat fluxes. Conversely, excessive impurity seeding can result in impurity accumulation in the core plasma, which degrades confinement and reduces overall fusion performance.

Traditional feedback control schemes for divertor detachment have been successfully demonstrated in multiple takamak devices \cite{TCV, DIII-D, KSTAR, EAST}. In these methods, impurity seeding is commonly used as the control actuator. In TCV, the poloidal location of the radiation front along the outer leg, reconstructed from the MANTIS diagnostic \cite{MANTIS}, serves as the control variable, while in DIII-D and KSTAR, the detachment degree, defined as a ratio of the measured ion saturation current, Jsat, by Langmuir probes (LPs) to the calculated Jsat from 2-pt model, is used as the control observer.

To enhance the performance and robustness of the detachment control algorithm for KSTAR, a surrogate model \cite{Zhu2,Zhu} has been developed based on a large dataset of UEDGE simulations. Approximately 70000 UEDGE steady states were generated, systematically scanning a wide range of edge plasma parameters, including upstream density, input power, impurity fraction, plasma current ($I_p$) and scaling of diffusivities. This massive simulation provides comprehensive coverage of the KSTAR operational space. Importantly, the UEDGE runs include self-consistent treatment of magnetic and electric drifts across the magnetic field (cross-field drfits), which are known to be critical for accurately capturing key detachment characteristics in the divertor such as strong in-out divertor asymmetries and strong effects of magnetic field (B-field) direction on detachment onset.

This paper is organized as follows. Section 2 introduces the setup of the UEDGE simulations and the generation of the database, including the parameter space scanned. Section 3 presents the key physics insights derived from the database, such as the role of divertor $T_e$ in detachment, the resulting detachment scaling, in–out divertor asymmetry, and plasma dynamics in response to actuator variations. Section 4 discusses potential applications of the database for detachment control. A summary is provided in Section 5. Appendix A describes the rationale behind the choice of the base model for the database, balancing accuracy and computational efficiency, while Appendix B outlines the workflow developed to overcome challenges in database production.

This database directly supports machine-learning surrogate models for predictive detachment control, bridging simulation and control applications. A first demonstration of detachment control using surrogate models trained on this database, including implementation of the same FOPDT dynamics for controller tuning, is reported separately~\cite{Zhu,GuptaArXiv2025}.


\section{Geomtry, deuterium plasma and impurity models}

The 2D edge plasma transport code UEDGE is used to generate the simulation database due to its robustness and flexibility in modeling divertor and SOL physics. UEDGE uses a fully implicit scheme for time integration, which provides strong numerical stability and enables the code to consistently converge in the presence of steep gradients and strong cross-field drifts. One of UEDGE’s key advantages for this work is its ability to quickly converge to a steady state when initialized from a state close to the final solution, making it especially well-suited for large-scale parameter scans where input conditions vary incrementally. This property allows for efficient simulations across the parameter space, significantly reducing the computational cost and time required to generate a comprehensive and physically consistent dataset. These features make UEDGE an ideal tool for constructing the high-fidelity simulation data for surrogate models.

\begin{figure}
  \centering
  \includegraphics[width=.45\textwidth]{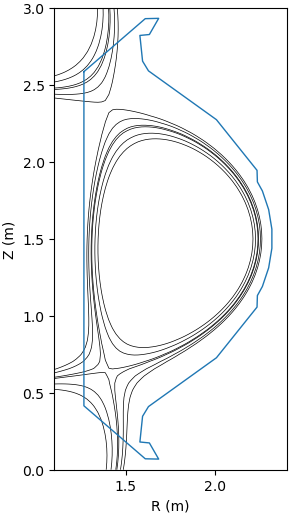}
  \caption{KSTAR equilibrium from the shot \#22849@56 is used to construct the base UEDGE case.}\label{Fig:equi}
\end{figure}

To initiate the large-scale database generation, a well-constructed base case is required as the starting point. The equilibrium used for the base case is from the KSTAR shot \#22849@56 with the carbon divertor, shown in Fig.~\ref{Fig:equi}.
The setup of the base case requires the balance between model accuracy and computational efficiency. After a careful analysis (see details in~\ref{appendix1}), the following setup is used:
\begin{itemize}
    \item \textbf{mesh resolution} A mesh with 64 poloidal cells and 24 radial cells is used, i.e. $nx \times ny = 64 \times 24$

\item \textbf{plasma transport models} Cross-field drifts are fully activated. Diffusivities for anomalous transport are specified as user-defined input profiles in UEDGE, and the details of how these profiles are constructed are provided in a later part of this section.

\item \textbf{plasma fueling} Deuterium ion density at the core boundary is fixed to maintain fueling.

\item \textbf{impurity transport models} Impurity transport models — The fixed-fraction carbon model, using a pre-tabulated radiation loss rates, is adopted as the default for computational efficiency, with validation against multi-charge-state runs confirming consistent detachment trends and plasma responses.

\end{itemize}

\begin{figure}
  \centering
  \includegraphics[width=.48\textwidth]{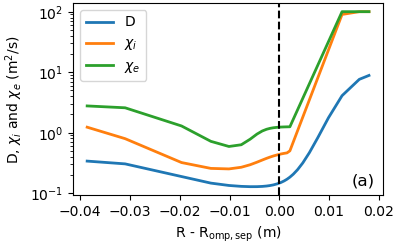}
  \hfill
  \includegraphics[width=.48\textwidth]{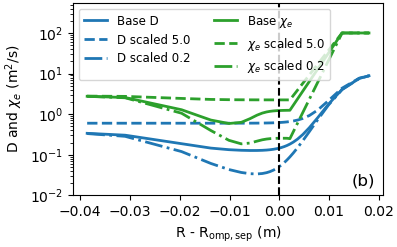}
  \caption{(a) Base profiles of D, $\chi_i$ and $\chi_e$ from SOLPS-ITER simulations; (b) Examples of applying scaling factors of $0.2$ and $5.0$ to the base profiles.}\label{Fig:DChi}
\end{figure}

In UEDGE, the anomalous transport coefficients, such as diffusivities for particle ($D_\perp$), ion heat ($\chi_i$) and electron heat ($\chi_e$), must be specified by the user as input profiles. The base profiles of $D_\perp$ and $\chi_i$ and $\chi_e$ are assumed by fitting the profiles used in SOLPS-ITER modeling of the KSTAR discharge \#22849@56, as shown in Fig.~\ref{Fig:DChi}. The base profiles are divided into 3 parts with 3 characteristic values: $D_\mathrm{core}$, $D_\mathrm{sep}$ and $D_\mathrm{SOL}$. The profiles are obtained by connecting $D_\mathrm{core}$, $D_\mathrm{sep}$ using a hyperbolic tangent function and connecting $D_\mathrm{sep}$ and $D_\mathrm{SOL}$ using an exponential function. A scaling factor $\alpha$ is applied to $D_\mathrm{sep}$ to consider transport uncertainties in real experiments. When $\alpha > 1.0$ and $D_\mathrm{core}$ is lower than the scaled $D_\mathrm{sep}$, $D_\mathrm{core}$ is replaced with the value of the scaled $D_\mathrm{sep}$. An example of applying factors of 0.2 and 5.0 to $D_\perp$ and $\chi_e$ are shown in Fig.~\ref{Fig:DChi}, the same for $\chi_i$ but not shown in the plot.

The database is generated by producing UEDGE steady-state solutions with scanning the uncertainty scaling factor $\alpha$ applied to the profiles of transport coefficients $D_\perp$, $\chi_e$ and $\chi_i$ used in UEDGE, and four important control parameters: 1) core boundary density for fueling, 2) input power, 3) carbon fraction, 4) plasma current. The scanned range for each control parameter is shown in Table~\ref{tab:table1}, which is meant to cover the operational space of KSTAR. The workflow used to generate the database, including the setup of initial conditions, sampling of control parameters, convergence criteria, and parallelization on HPCs, is described in~\ref{appendix2}.
\begin{table}[]
    \centering
    \begin{tabular}{|c|c|c|c|c|c|}
        \hline
        Control parameters & Range & Number of data points \\ \hline
        Density & 1 - 8 & 30 \\ \hline
        Power & 1 - 9 MW & 15 \\ \hline
        C fraction & 0\% - 6\% & 10 \\ \hline
        Ip & 300 - 800 kA & 5 \\ \hline
        D scaling & 0.6 - 2.0 & 10 \\ \hline
    \end{tabular}
    \caption{Ranges of the scanned control parameters: density at the core boundary ranging from $1 - 8\times 10^{19}\,\mathrm{m}^{-3}$, input power ranging from $1-9\,\mathrm{MW}$, carbon fraction ranging from $0-6\%$, plasma current ranging from $300 - 800\,\mathrm{kA}$, and diffusivity scaling factor ranging from $0.6 - 2.0$. The numbers of sampled points for each scanned parameter are listed in the third column.}
    \label{tab:table1}
\end{table}

\section{Characteristics of KSTAR detachment physics}
\label{sec:steady_database}

Various detachment characteristics can be investigated using the UEDGE database. In this section, several key aspects are examined. Section 3.1 explores the role of electron temperature at the target plate in detachment, following the methodology outlined in~\cite{Stangeby2018}. In Section 3.2, the correlation between electron density, impurity fraction, and power input at the onset of detachment (i.e., detachment scaling) is derived from the UEDGE database. Section 3.3 focuses on the in–out divertor asymmetry of KSTAR. Finally, Section 4.4 analyzes the plasma response to control actuators, such as gas puffing.

The discussions throughout this section are centered on the concept of divertor detachment. Here, partial and complete detachment are not explicitly distinguished, as the roll-over of the ion saturation current density ($j_\mathrm{sat}$) at the outer strike point (OSP) and the roll-over of the total ion saturation current ($I_\mathrm{sat}$) integrated over the outer divertor plate occur at similar collisionality levels. Therefore, within a given density scan—where input power, impurity fraction, transport coefficients, and plasma current are fixed—the case exhibiting the maximum $j_\mathrm{sat}$ is identified as the point of 'detachment onset'.

\subsection{Role of $T_\mathrm{et}$ in detachment }

\subsubsection{$T_\mathrm{et}$ and Radiation front at detachment onset.}

\begin{figure}
  \centering
  \includegraphics[width=.5\textwidth]{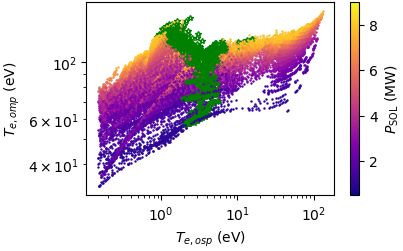}
  \caption{Outer midplane separatrix electron temperature $T_\mathrm{e,omp}$ as a function of outer strike point electron temperature $T_\mathrm{e,osp}$ for all cases converged with $I_p = 700\,\mathrm{kA}$. The cases at outer divertor detachment onset are colored in green. Colormap represents input power.}
  \label{Fig:onset}
\end{figure}

Fig.~\ref{Fig:onset} shows the outer mid-plane electron temperature $T_\mathrm{e,omp}$ at the separatrix as a function of outer strike-point electron temperature $T_\mathrm{e,osp}$.
$T_\mathrm{e,omp}$ is mostly determined by the input power, which is consistent with the 2-pt model. $T_\mathrm{e,osp}$ at detachment onset on the outer target (colored in green) are within $\sim 1-5 \,\mathrm{eV}$, with most cases concentrating $\sim 3-4\,\mathrm{eV}$.

\begin{figure}
  \centering
  \includegraphics[width=.48\textwidth]{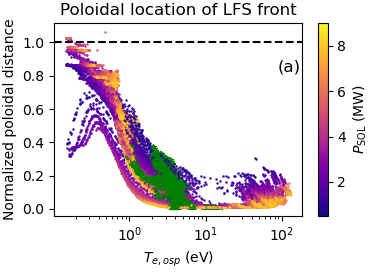}
  \hfill
  \includegraphics[width=.48\textwidth]{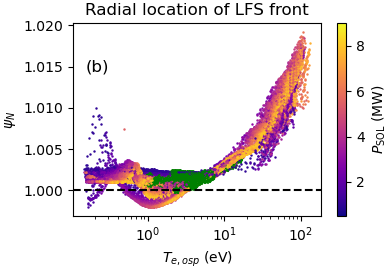}
  \caption{(a) The poloidal location of the LFS radiation front as a function of $T_\mathrm{e,osp}$ for all cases with $I_p = 700\,\mathrm{kA}$. The distance away from the outer target plate is normalized by the poloidal length of the outer leg. The dashed line denotes the location of the X-point. (b) The radial location of the LFS radiation front as a function of $T_\mathrm{e,osp}$. The dashed line denotes the location of the separatrix ($\psi_N = 1.0$). For both plots, detachment onset cases are colored in green and colormap represents input power.} \label{Fig:front}
\end{figure}
It can be inferred, shown in Fig.~\ref{Fig:front}, that the LFS radiation front is highly correlated with the $T_\mathrm{e,osp}$. The strong radiation is generally near the outer target plate and can extend radially outward from the separatrix location (OSP), $\psi_n \sim 1.02$, when $T_\mathrm{e,osp} \sim 100\,\mathrm{eV}$. As $T_\mathrm{e,osp}$ decreases, either by increasing carbon fraction or decreasing input power, the radiation front first moves towards the OSP along the target plate. Detachment onset occurs when the radiation front moved to the OSP and is about to/already detach from the target. As the front moves away from the plate along the outer leg, the outer divertor evolves to a deep detached state.

Consistent with~\cite{Stangeby2018}, outer target electron temperature $T_e$ is a robust indicator of detachment onset, occurring at $T_\mathrm{e,osp} \sim 3-4\,\mathrm{eV}$, largely insensitive to input power, upstream density, carbon fraction, or diffusivities.

\begin{figure}
  \centering
  \includegraphics[width=\textwidth]{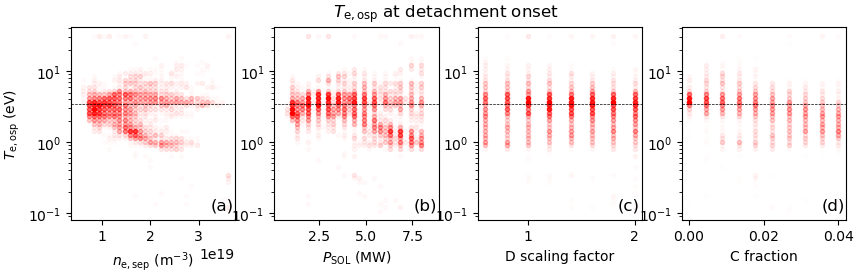}
  \caption{Outer strike point electron temperature $T_\mathrm{e,osp}$ at detachment onset as a function of outer midplane separatrix electron density $n_\mathrm{e,sep}$ (a), power flowing through the separatrix $P_\mathrm{SOL}$ (b), D scaling factor (c), and carbon fraction (d). The darkness represents the number of the converged cases. All the cases shown are with $I_p = 700\,\mathrm{kA}$. The temperature of $3.5\,\mathrm{eV}$ is denoted as the dashed line, around which most of the cases are located.}\label{Fig:Teosp_onset}
\end{figure}

\subsubsection{Correlation of $T_\mathrm{et}$ and momentum and power loss.}

\begin{figure}
  \centering
  \includegraphics[width=\textwidth]{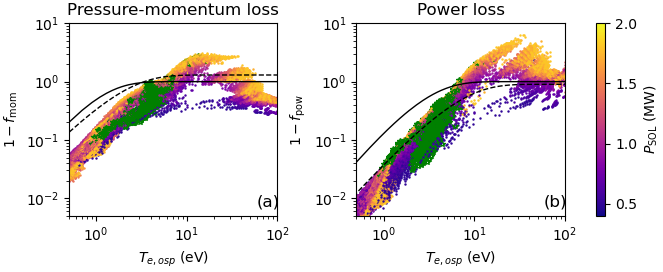}
  \caption{Presurre-momentum loss $1-f_\mathrm{mom}$ as a function of outer strike point electron temperature $T_\mathrm{e,osp}$ (a); Power loss $1-f_\mathrm{mom}$ as a function of outer strike point electron temperature $T_\mathrm{e,osp}$ (b). Dots: UEDGE simulations with $I_p = 700\,\mathrm{kA}$. Cases at detachment onset are colored in green. 
  Black curves: scaling $1-f_\mathrm{momloss} = (1-e^{-T_\mathrm{e,osp}/0.8})^{2.1}$ (solid in (a)), scaling $1-f_\mathrm{momloss} = 1.3(1-e^{-T_\mathrm{e,osp}/1.8})^{1.6}$ (dashed in (a)), $1-f_\mathrm{powloss} = (1-e^{-T_\mathrm{e,osp}/2.4})^{1.9}$ (solid in (b)), $1-f_\mathrm{momloss} = 0.9(1-e^{-T_\mathrm{e,osp}/6.0})^{1.7}$ (dashed in (b)) taken from~\cite{Stangeby2018}. Colormap represents input power.}\label{Fig:fmom}
\end{figure}

\begin{figure}
  \centering
  \includegraphics[width=0.5\textwidth]{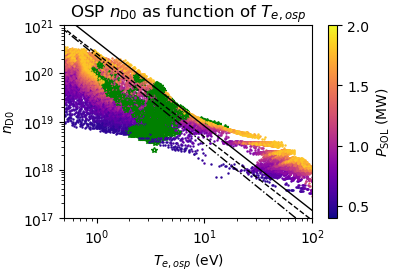}
  \caption{Neutral density ($D_0$ in UEDGE and $D_2$ in SOLPS-ITER and the scaling taken from~\cite{Stangeby2018}) as a function of outer strike point electron temperature $T_\mathrm{e,osp}$. Dots: UEDGE simulations with $I_p = 700\,\mathrm{kA}$. Cases at detachment onset are colored in green. 
  Black curves: scaling $T_\mathrm{e,osp} = 5.99\times 10^{11}n_{D_2}^{-0.57}$ (solid) scaling $T_\mathrm{e,osp} = 6.01\times 10^{11}n_{D_2}^{-0.577}$ (dashed) scaling $T_\mathrm{e,osp} = 1.57\times 10^{11}n_{D_2}^{-0.55}$ (dotted dashed) taken from~\cite{Stangeby2018} where it is claimed that these scalings only apply to when the target $T_e < 10\,\mathrm{eV}$. Colormap represents input power.}\label{Fig:nDt}
\end{figure}

The relative insensitivity of target $T_e$ to upstream conditions near detachment onset arises from the strong nonlinear dependence of atomic processes on $T_e$, which become dominant when $T_e$ drops to $\sim$ a few eV. These processes, e.g. radiation and charge-exchange, lead to enhanced volumetric power loss and pressure-momentum loss along flux tubes between the divertor entrance and target, both of which are essential for achieving detachment. Focusing on the flux tube adjacent to the separatrix in the SOL and following~\cite{Stangeby2018}, the pressure-momentum loss and volumetric power loss along the flux tube are defined as:
\begin{align}
    1-f_\mathrm{momloss} &= \frac{p_\mathrm{tot}^\mathrm{t}}{p_\mathrm{tot}^\mathrm{X}} \label{equ:fmom} \\
    1-f_\mathrm{powloss} &= \frac{q_\parallel^\mathrm{t}A^\mathrm{t}b_\mathrm{x}^\mathrm{t}}{q_\parallel^\mathrm{X}A^\mathrm{X}b_\mathrm{x}^\mathrm{X}} \label{equ:fpow}
\end{align}
where the variables with superscript t and X denote the variables at the OSP and the outer divertor entrance near the X-point. $A$ is the poloidal cross-section area of the flux tube, $b_x$ is the magnetic pitch angle and $q_\parallel$ is the parallel heat flux along the magnetic field line including both electron and ion contribution. The total pressure is defined as the sum of electron pressure, ion pressure and ion drifting dynamic pressure:
\begin{align}
    p_\mathrm{tot} = n_eT_e+n_iT_i+n_im_iu_{i\parallel}^2
\end{align}
The scalings of pressure-momentum and volumetric power losses based on the UEDGE database with $I_p = 700\,\mathrm{kA}$ are derived:
\begin{align}
    1-f_\mathrm{momloss}=0.942(1-e^{-T_\mathrm{e,osp}/1.933})^{2.656} 
\end{align}
shown in Fig.~\ref{Fig:fmom}. The volumetric power loss becomes increasingly significant as $T_\mathrm{e,osp}$ drops below $\sim 10\,\mathrm{eV}$ whereas the pressure-momentum loss begins around $T_\mathrm{e,osp} \sim 2-3 \,\mathrm{eV}$, consistent with the SOLPS-ITER modeling of the same equilibrium with various input power and various levels of gas puff (shown as stars in Figs.~\ref{Fig:fmom} and~\ref{Fig:nDt}), and with the scaling of pressure-momentum and power losses derived based on various SOLPS-ITER simulations of other devices (shown as solid and dashed curves)~\cite{Stangeby2018}. The onset of detachment roughly coincides with the beginning of significant pressure-momentum loss.

However, UEDGE tends to predict stronger pressure-momentum losses at $T_\mathrm{e,osp} < 1 \,\mathrm{eV}$, which is likely attributable to its fluid approximation of neutrals ignoring atom transport between the outer-most flux tube of the plasma simulation and the actual vessel wall, and the omission of molecular effects. In this database, molecules are not used, so, the only neutral species is the deuterium atom.  Recycled neutrals are assumed to be atomic only, which can lead to an overestimation of atom density especially at $T_\mathrm{e,osp} \sim 1\,\mathrm{eV}$ when thermal emmission of molecules as recycled become more significant. As a result, charge-exchange collisions — responsible for much of the plasma pressure loss — may be overestimated. It can be seen that the atom density predicted by UEDGE is comparable to molecule density in SOLPS-ITER modeling and the scaling of molecule density derived from various SOLPS-ITER simulations of other devices~\cite{Stangeby2018}, shown in Fig.~\ref{Fig:nDt}.

\subsection{Detachment scaling laws}

\begin{figure}
  \includegraphics[width=.49\textwidth]{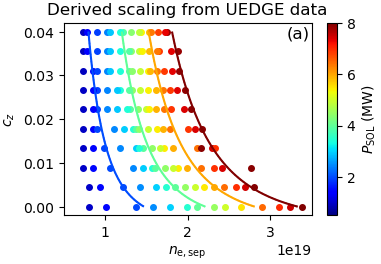}
  \hfill
  \includegraphics[width=.49\textwidth]{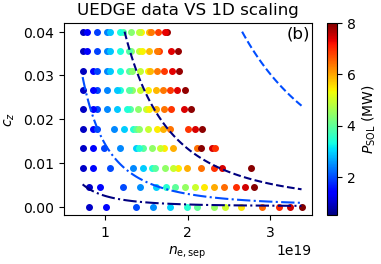}
  \caption{Carbon fraction $c_z$ as a function of outer midplane separatrix electron density $n_\mathrm{e,sep}$ for various input power, for all cases with $I_p = 700\,\mathrm{kA}$ and D scaling factor of $1.0$. (a) UEDGE (dots) versus derived scaling law (solid curves) from Eq.~\ref{equ:detachment_scaling}; (b) UEDGE (dots) versus scaling derived from 1D HERMES simulations (dashed curves)~\cite{Body} and the rescaled 1D scaling, with a factor of 0.2 applied to $P_\mathrm{SOL}$ (dotted dashed curves). Colormap represents input power.}\label{Fig:dets}
\end{figure}

Existing scaling laws for detachment onset derived theretically~\cite{Goldston} or from 1D edge modeling~\cite{Body} are used to guide divertor designs or divertor experiments. Similarly, from this 2D UEDGE database for a certain $I_p$ and D scaling factor, a detachment scaling law for power across the separatrix  as a function of separatrix electron density at the outer midplane $n_\mathrm{e,sep}$ and carbon fraction $c_z$ at detachment onset can be derived from the database, shown in Fig.~\ref{Fig:dets}:
\begin{align}
    \left(\frac{n_\mathrm{e,sep}}{10^{20}\,\mathrm{m}^{-3}}\right)^2\left(\frac{c_z+0.432\%}{1\%}\right)^{0.524} = 0.00615 \left(\frac{P_\mathrm{SOL}}{1 \,\mathrm{MW}}\right)^{1.180}\label{equ:detachment_scaling}
\end{align}
As a comparison, the scaling derived based on 1D HERMES simulations using fixed-fraction neon~\cite{Body} is also shown. The original 1D scaling uses $q_\parallel$ at upstream instead of input power. Here, the heat flux is calculated by $q_\parallel = P_\mathrm{SOL}/\lambda_q$ where $\lambda_q \sim $ from Eich's scaling~\cite{Eich}. This overestimates the $q_\parallel$, so, for the same power input, 1D scaling overestimate $n_\mathrm{e,sep}$ or $c_z$ required to achieve detachment. A scaling factor of 0.2 is applied to $P_\mathrm{SOL}$ in the 1D scaling. After this rescaling, the primary difference between the UEDGE-based scaling and the 1D scaling lies in their dependence on impurity fraction. Although different impurity species are used, the 1D scaling tends to overestimate the effect of impurities; specifically, it predicts that increasing impurity concentration causes $n_\mathrm{e,sep}$ required for detachment onset to decrease more rapidly than predicted by the UEDGE scaling (i.e., a steeper slope).
\begin{figure}
\centering  \includegraphics[width=.49\textwidth]{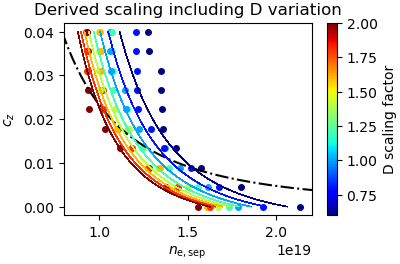}
  \caption{Carbon fraction $c_z$ as a function of outer midplane separatrix electron density $n_\mathrm{e,sep}$ for various D scaling factors, for all cases with $I_p = 700\,\mathrm{kA}$ and $P_\mathrm{SOL} = 3\,\mathrm{MW}$. Dots: UEDGE simulations; Solid curves: derived scaling law as Eq.~\ref{equ:scaling_dif}; Dotted dashed curves: rescaled 1D scaling with a factor of 0.2 applied to $P_\mathrm{SOL}$. Colormap represents D scaling factor.}\label{Fig:dets_dif}
\end{figure}

The diffusivity has strong effects in detachment~\cite{XuPSI2025}. Along a certain flux tube (from 1D point of view), it is the parallel heat flux that matters for detachment. $q_\parallel \propto P_\mathrm{SOL}/\lambda_q$ and $\lambda_q$ is determined by the competition between the parallel transport along the field line and radial transport due to cross-field drifts and turbulence represented by anomalous transport using $D$ and $\chi$ in UEDGE. Consider a radial diffusive process (here the effects of $D$ and $\chi$ are not separated in this simple picture):
\begin{align}
    \frac{\partial n}{\partial t} &= - D \nabla^2_r n \\
    u_r &= -D \frac{\nabla_r n}{n} = \frac{D}{\lambda} \\
    \lambda &= u_r\delta t = \frac{D\delta t}{\lambda} \\
    \lambda &\propto \sqrt{D\delta t}\label{equ:coef0.5}
\end{align}
where $\delta t$ is parallel transport time from upstream to the target plate, determined by ion parallel velocity in the order of acoustic speed, $\delta t\propto 1/\sqrt{T_e}$. Therefore, $D$ should enter the denominator of the term with $P_\mathrm{sol}$. However, a large fraction of the radial transport is due to cross-field drifts that can be treated using a critical diffusivity coefficient $D_\mathrm{crit}$, following the study in~\cite{XuNF2019} where it shows the total radial transport $D_\mathrm{tot}$ is not simply a sum of $D_\mathrm{crit}$ and $D$. The total transport coefficient $D_\mathrm{tot}$ is in the order of $\sim D_\mathrm{crit}$ when the anomalous diffusivity $D$ is lower than $D_\mathrm{crit}$. Then $D_\mathrm{tot}$ is in the order of $\sim D$ when $D$ is higher than $D_\mathrm{crit}$. The flux limiter concept is borrowed to construct the total $D_\mathrm{tot}$ that fits the derived form in~\cite{XuNF2019}:
\begin{align}
    D_\mathrm{tot} = \frac{D-D_\mathrm{crit}}{1+(D_\mathrm{crit}/D)^3} + D_\mathrm{crit}
\end{align}
A new scaling law including the variation of D/Chi is derived:
\begin{align}
    \left(\frac{n_\mathrm{e,sep}}{10^{20}\,\mathrm{m}^{-3}}\right)^2\left(\frac{c_z+0.558\%}{1\%}\right)^{0.506} = 0.00155 \frac{(P_\mathrm{SOL}/1 \,\mathrm{MW})^{1.50}}{D_\mathrm{tot}^{0.520}}\label{equ:scaling_dif}
\end{align}
with
\begin{align}
    D_\mathrm{crit} = 0.0435
\end{align}
The comparision between the UEDGE data and the scaling derived for $P = 3\,\mathrm{MW}$ and $I_p = 700\,\mathrm{kA}$ can be seen in Fig.~\ref{Fig:dets_dif}. The scaling power to the $n_\mathrm{e,sep}$ and $c_z$ are somewhat unchanged as expected. The power to $P_\mathrm{SOL}$ is increased mainly due to the dependence of $\delta t$, which is merged into the numerator, on upstream electron temperature which is determined mainly by $P_\mathrm{SOL}$. The power to $D_\mathrm{tot}$ derived is $0.52$, close to the prediction by Eq.~\ref{equ:coef0.5}.

\begin{figure}
  \centering
  \includegraphics[width=.49\textwidth]{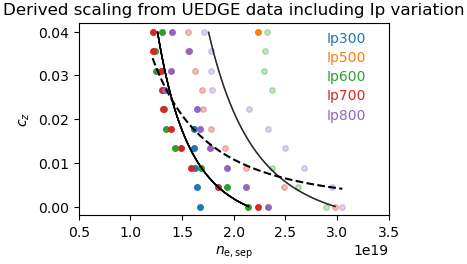}
  \caption{Carbon fraction $c_z$ as a function of outer midplane separatrix electron density $n_\mathrm{e,sep}$ for various D scaling factors, for all cases with $P_\mathrm{SOL} = 3\,\mathrm{MW}$ and D scaling factors of $0.6$ (lighter color) and $2.0$ (solid color). Dots (UEDGE simulations): \textcolor{C0}{blue} $I_p = 300\,\mathrm{kA}$,  \textcolor{C1}{orange} $I_p = 500\,\mathrm{kA}$, 
\textcolor{C2}{green} $I_p = 600\,\mathrm{kA}$, \textcolor{C3}{red} $I_p = 700\,\mathrm{kA}$, \textcolor{C4}{purple} $I_p = 800\,\mathrm{kA}$; Curves: derived scaling law as Eq.~\ref{equ:scaling_dif} using diffusivity with D scaling factor of $2.0$ (solid black) and $0.6$ (lighter solid black), rescaled 1D scaling with a factor of 0.2 applied to $P_\mathrm{SOL}$ (dashed black).}\label{Fig:dets_Ip}
\end{figure}

The database includes five different equilibria with varying plasma currents. However, the convergence rates for the lower-current cases ($300\,\mathrm{kA}$ and $500\,\mathrm{kA}$) are relatively poor and do not exhibit the clear trends observed at higher plasma currents, indicating reduced data quality in this regime. Consequently, the available variations in plasma current are insufficient to robustly derive a scaling law that incorporates plasma current dependence. With the limited lower-current data, plasma current effects are observable at lower diffusivity (lighter colors in Fig.~\ref{Fig:dets_Ip}), but become negligible at higher diffusivity (darker colors in Fig.~\ref{Fig:dets_Ip}).

%

\noindent
The scaling in Eq.~(12) captures the empirical dependence of the detachment threshold on
the input power and impurity concentration, consistent with experimental observations
from AUG and KSTAR, where increasing $P_{\mathrm{SOL}}$ or reduced impurity fraction delays
detachment~\cite{Henderson2019NME,Henderson2023NF}.
The relatively weak impurity sensitivity obtained from the UEDGE database agrees with
the earlier comparison presented in this section~\cite{Body}, which showed that
one-dimensional HERMES models tend to overestimate the impact of impurities on
$n_{e,\mathrm{sep}}$.
Thus, Eq.~(12) reproduces the general functional dependencies found in analytical
Lengyel--Goedheer and HERMES regressions but with a flatter slope in $c_z$, indicating
a reduced impurity leverage in two-dimensional transport.
Comparable power and seeding trends have also been observed in DIII--D detachment
control experiments~\cite{Wang2023NF}, further supporting the robustness of the
UEDGE-based scaling.

\subsubsection*{\bf Exploratory normalization and extrapolation}
To compare with empirical density limits and assess behavior at higher power, 
the detachment threshold is reformulated in Greenwald-normalized form, 
$n_{\mathrm{th,GW}} = n_{e,\mathrm{sep}} / n_\mathrm{GW}$, where $n_{GW} = I_p / (\pi a^2)$ and $n_{\mathrm{e,line-ave}}$ is line-averaged density, 
to provide orientation beyond the KSTAR operating domain. 
This normalization reduces residual dependence on $I_p$ and $a$ within the 
KSTAR-scale database, but its extrapolation to reactor conditions remains 
exploratory, as the present dataset does not yet encompass reactor-scale size, 
current, or ITER-specific magnetic configurations, divertor conditions, and 
plasma parameters.
\begin{equation}
\label{equ:scaling_norm_G}
n_{\mathrm{th,{GW}}} =
\frac{n_{\mathrm{e,sep}}}{n_{\mathrm{GW}}}
= 
\left[
0.00155\,
\frac{(P_{\mathrm{SOL}}/1\,\mathrm{MW})^{1.50}}
     {D_{\mathrm{tot}}^{0.520}
      \left(\dfrac{c_z + 0.558\%}{1\%}\right)^{0.506}}
\right]^{1/2}
\left(\frac{\pi a^2}{I_p}\right).
\end{equation}
with
\begin{equation}
\label{equ:Dcrit}
D_{crit} = 0.0435 \mathrm{m^2/s}.
\end{equation}
\noindent
Here, $n_{\mathrm{th,GW}} = n_{e,\mathrm{sep}}/n_{\mathrm{GW}}$ denotes the separatrix
density normalized to the Greenwald limit, providing a cross-machine measure of
detachment accessibility. This normalization follows the theoretical framework
proposed by Goldston~\cite{Goldston}, which highlights the importance of the normalized density and the radiative power balance in
setting the exhaust limit. Equation~(14) adopts this normalized form, enabling
consistent comparison and extrapolation to reactor-scale conditions.
Recasting Eq.~(14) using
$P_{\mathrm{SOL}} = q_{\parallel}(2\pi R,\lambda_q)(B_p/B)$
together with $\lambda_q \propto B_p^{-\gamma}$ introduces explicit $B$ and $R$ dependences that are qualitatively consistent with Reinke’s interpretation~\cite{Reinke2017NF}: higher $B$ leads to larger $q_{\parallel}$ and thus makes detachment more difficult, whereas larger $R$—through a longer connection length $L_{\parallel}$—enhances detachment accessibility. The net dependence ultimately depends on whether the comparison is made at fixed $P_{\mathrm{SOL}}$ or fixed $q_{\parallel}$.

Solving the normalized fit [Eq.~(\ref{equ:scaling_norm_G})] for $n_{\mathrm{th,GW}} = 0.6$, assuming the ratio of OMP separatrix density to line-averaged density $n_\mathrm{e,sep}/n_\mathrm{e,line-ave}\sim 0.6$, 
yields the impurity fraction needed to achieve detachment onset as a function of input power $P_{\mathrm{SOL}}^{(\bar n=1)}$ at line-averaged density reaching the Greenwald density. For typical KSTAR parameters ($R = 1.8$~m, $a = 0.5$~m, 
$B_t \sim 2$~T, $I_p \sim 0.8$~MA), $n_{GW} \simeq 1.0\times10^{20}\,\mathrm{m}^{-3}$. Taking $D_\mathrm{tot} = D_\mathrm{crit} = 0.0435$ $m^2/s$, and 
assuming impurity concentrations $\lesssim 10\%$, this yields
$P_{\mathrm{SOL}}^{(\bar n=1)} \approx 28$–$30$~MW, which represents the approximate power threshold below which detachment is expected to be achievable. When evaluated for the default database parameters
($D_{\mathrm{scaling}} = 2.0$ $m^2/s$, $f_z = 1\%$), the same relation yields
$P_{\mathrm{SOL}}^{(\bar{n}=1)} \approx 13$–$14~\mathrm{MW}$ (with a 95\% confidence
interval of $\pm 3~\mathrm{MW}$), which lies within the experimentally accessible
range on KSTAR and provides a practical, verifiable benchmark for approaching the
empirical density limit in future high-power detachment studies.
\begin{figure}
\centering  
\includegraphics[width=0.5\textwidth]{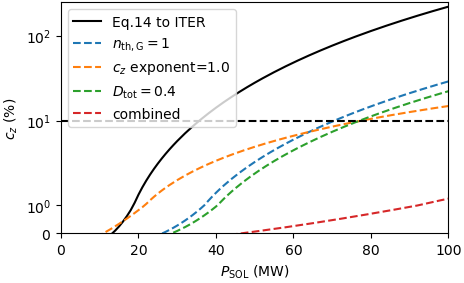}
\caption{Impurity fraction required to achieve detachment onset as a function of $P_{\mathrm{SOL}}$ based on Eq.~\ref{equ:scaling_norm_G}. The solid curve shows the KSTAR fit, Eq.~\ref{equ:scaling_norm_G} assuming $n_\mathrm{th,G} = 0.6$ (line-averaged density $\sim$ the Greenwald density), $D_\mathrm{tot} = 0.0435$, directly applied to the ITER parameters. Dashed lines: (\textcolor{C0}{blue}) assuming $n_\mathrm{th,G} = 1$; (\textcolor{C1}{orange}) replacing the exponent ($0.506$) on $c_z$ in Eq.~\ref{equ:scaling_norm_G} with $1.0$ to mimic radiation due to higher-Z impurities; (\textcolor{C2}{green}) assuming $D_\mathrm{tot} = 0.4$ for enhanced turbulence; (\textcolor{C3}{red}) combining all the effects.}
\label{fig:normalized_nsep_threshold}
\end{figure}

Although the empirical scaling is derived from KSTAR-scale simulations, it is
instructive to examine how the same functional form behaves when applied to
reactor-scale parameters. For ITER-like conditions
($R = 6.2~\mathrm{m}$, $a = 2.0~\mathrm{m}$, $B_t = 5.3~\mathrm{T}$,
$I_p = 15~\mathrm{MA}$, $n_{\mathrm{GW}} \simeq 1.2\times10^{20}~\mathrm{m^{-3}}$),
a direct extrapolation of the KSTAR-based detachment scaling would predict that achieving detachment onset at $P_{\mathrm{SOL}} > 70\text{--}80~\mathrm{MW}$ requires an unphysical impurity fraction exceeding $100\%$. This underscores that direct application of the KSTAR fit outside its validated parameter space is not physically meaningful, even when the line-averaged density approaches the Greenwald limit. This apparent overshoot highlights that direct application of the KSTAR scaling outside its database domain is not physically meaningful, but it does indicate the scale at which additional physics must be considered. However, comprehensive SOLPS–ITER simulations of the $Q = 10$ baseline scenario~\cite{Pitts2013JNM,Stangeby2022NF,Pitts2019NME} predict detachment onset near $n_{e,\mathrm{sep}} \simeq (0.8\text{--}1.0)\times10^{20}~\mathrm{m^{-3}}$ ($n_{\mathrm{th,GW}} \approx 1$) when approximately $70\%$ of the input power is radiated. Accordingly, a power range of $P_{\mathrm{SOL}} \approx 100$–$150~\mathrm{MW}$ represents the regime in which ITER is designed to sustain detachment close to the Greenwald limit through strong radiative losses rather than a quantitative prediction.

To understand the source of these strong radiative losses, it is useful to consider the role of impurity radiation in modifying the scaling behavior. The original KSTAR scaling is based on low-$Z$ carbon impurity cases, which yield an exponent of approximately $0.5$ on impurity fraction. Higher-$Z$ impurity radiation, which is needed for ITER, is expected to increase this exponent, e.g. a value of $\sim 0.72$ for nitrogen in ASDEX-U and JET~\cite{Henderson2022NME}. Therefore, a comparison is made using a theoretical exponent of $1.0$ (following Goldston), representative of higher-$Z$ seeded-impurity operation (see orange curve in
Fig.~11).

Regarding the density limit, recent DIII–D experiments have demonstrated that the line-averaged density can exceed the Greenwald value ($f_{\mathrm{Gr}} > 1$) in high–$\beta_P$ regimes when the pedestal density remains below the limit ($f_{\mathrm{Gr,ped}} \lesssim 0.7$–$0.8$), supported by the formation of an internal transport barrier that enhances core peaking and suppresses edge turbulence~\cite{Ding2024Nature}.
In particular, the recent negative-triangularity campaign on DIII–D achieved line-averaged densities up to twice the Greenwald value~\cite{Thome2024}, while the separatrix density remained capped near $n_{\mathrm{GW}}$. This suggests that $n_{e,\mathrm{sep}}/n_{\mathrm{GW}}$ is a more fundamental measure than $n_{e,\mathrm{line}}/n_{\mathrm{GW}}$ for assessing the density limit and detachment accessibility. Motivated by this observation, $n_\mathrm{th,G} = 1$ ($n_\mathrm{e,sep}\sim n_\mathrm{GW}$) is applied to the density limit in the scaling (see blue curve in Fig.~\ref{fig:normalized_nsep_threshold}) for comparison with the original KSTAR fit.

Finally, both BOUT++~\cite{ZLi2019,Nami2020} and XGC~\cite{Chang2017} simulations indicate that the effective cross-field diffusivity driven by turbulence can significantly exceed the nominal critical value used in the scaling formulation. A factor of $\sim 10$ increase in effective $D_{\mathrm{tot}}$ is tested in the scaling (see green curve in Fig.~\ref{fig:normalized_nsep_threshold}), yielding an effect comparable to applying a factor of two to the Greenwald density limit. The scanned range of effective diffusivity in the KSTAR database remains relatively narrow—about a factor of four—but extending this range could increase the fitted exponent on $D_{\mathrm{tot}}$ (from $\sim 0.25$ in $D_{\mathrm{tot}}^{0.251}$), implying a stronger dependence of the detachment threshold density, $n_{e,\mathrm{sep}}/n_{\mathrm{GW}}$, on turbulent transport. By combining the effects of (1) higher-$Z$ impurity radiation efficiency, (2) enhanced turbulent transport, and (3) Greenwald-limit extension due to core-peaked density profiles, the modified scaling suggests that achieving detachment in ITER-like regimes becomes substantially more accessible (see red curve in Fig.~\ref{fig:normalized_nsep_threshold}).

\subsection{Comparison with Eich’s $\lambda_q$–$B_p$ scaling}
\begin{figure}
\centering  \includegraphics[width=.49\textwidth]{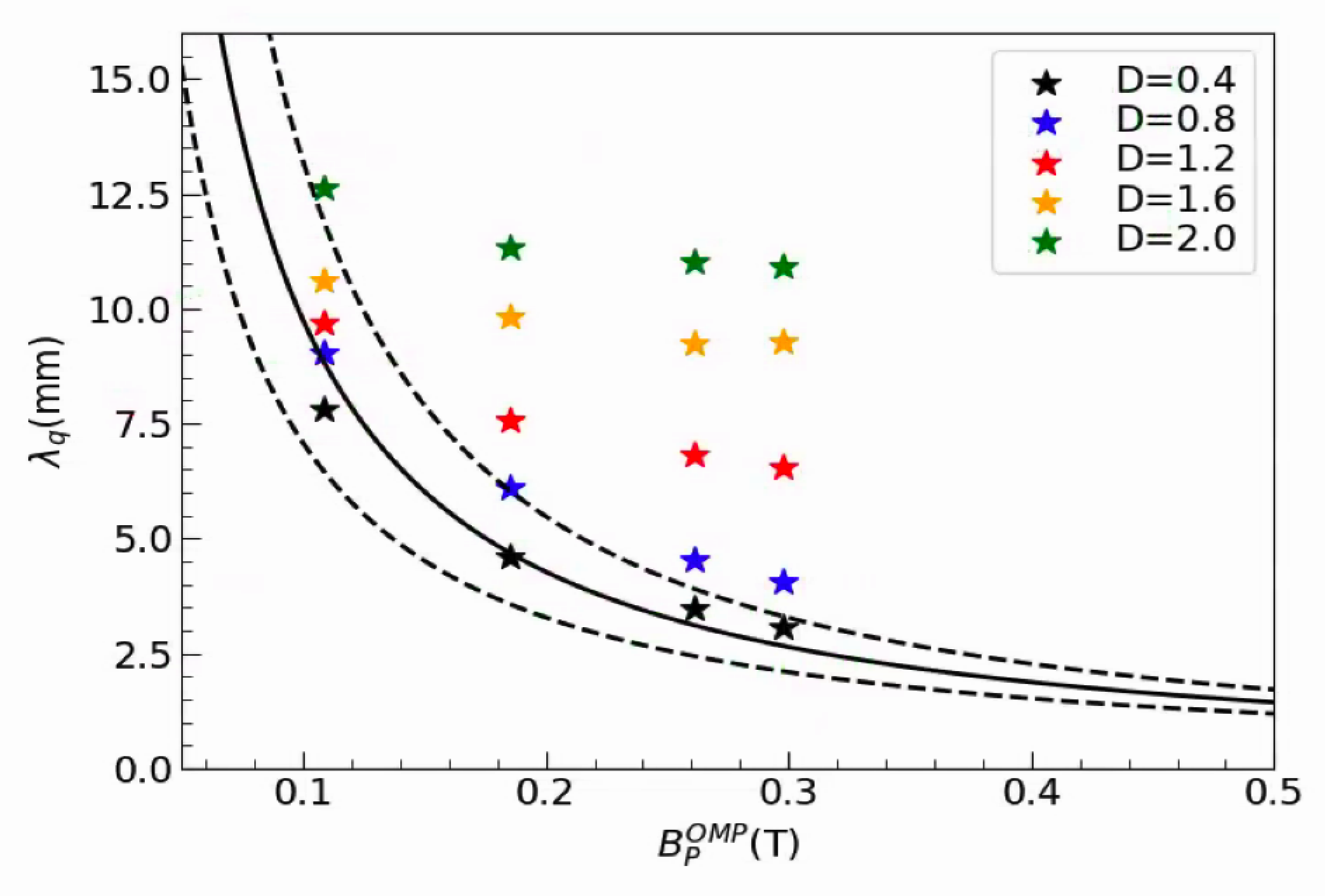}
\caption{Heat–flux width $\lambda_q$ versus poloidal magnetic field $B_p$ (evaluated at the outboard midplane separatrix). Solid curve: Eich multi-machine scaling~\cite{Eich}; dashed lines: $\pm1\sigma$ scatter of the fit. Stars (UEDGE, spatially uniform transport): \textcolor{black}{black} $(D,\chi)=(0.4,1.0)$, \textcolor{blue}{blue} $(0.8,2.0)$, \textcolor{orange}{orange} $(1.6,4.0)$, \textcolor{green!70!black}{green} $(2.0,5.0)$~$\mathrm{m^2/s}$.}
\label{fig:lambdaq-bp}
\end{figure}

At low collisionality and across equilibria with varying $I_p$, UEDGE heat–flux widths computed with the \emph{radially varying} $D/\chi$ from Fig.~\ref{Fig:DChi} deviate from Eich’s multi-machine $\lambda_q$–$B_p$ scaling~\cite{Eich}. A likely driver is the large imposed radial variation in transport (up to $\times10$ in $D$ and $\times100$ in $\chi$). To isolate this, we reran representative cases with \emph{spatially uniform} transport, $(D,\chi)\in\{(0.4,1.0),(0.8,2.0),(1.6,4.0),(2.0,5.0)\}$~$\mathrm{m^2/s}$. 
As shown in Fig.~\ref{fig:lambdaq-bp}, lower $(D,\chi)$ sets align with Eich’s trend and scatter, whereas higher $(D,\chi)$ produce systematically broader $\lambda_q$, demonstrating the sensitivity of $\lambda_q$ to the assumed radial transport profiles.

\subsection{Characteristics of in-out asymmetry}
The database shows a strong in-out asymmetry between the inner and outer plasma states. For most cases, $T_\mathrm{e,osp}$ in the outer divertor is much lower than $T_\mathrm{e,isp}$ in the inner divertor. This characteristic of in-out asymmetry in KSTAR is quite different from observed in ASDEX-U~\cite{HFHD}, JET~\cite{JET2015}, and DIII-D~\cite{JarvinenNME2017,MaNF2021,MaurizioNF2024,ScottiNT} where the outer divertor is more attached than the inner divertor with ion $\nabla B$ into the divertor, due to ExB drifts driving particles from the outer divertor into the inner divertor.

\subsubsection{In-out asymmetry in divertor detachment.}
\begin{figure}
  \centering
  \includegraphics[width=.5\textwidth]{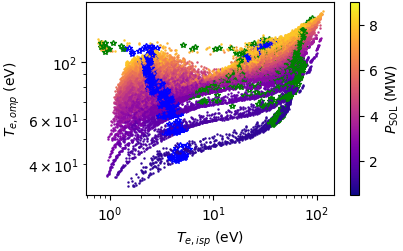}
  \caption{Outer midplane separatrix electron temperature $T_\mathrm{e,omp}$ as a function of inner strike point electron temperature $T_\mathrm{e,isp}$ for all cases converged with $I_p = 700\,\mathrm{kA}$. The cases at outer divertor detachment onset are colored in green and those at inner divertor detachment onset are colored in blue. Colormap represents input power.}\label{Fig:onset_inner}
\end{figure}

\begin{figure}
  \centering
  \includegraphics[width=\textwidth]{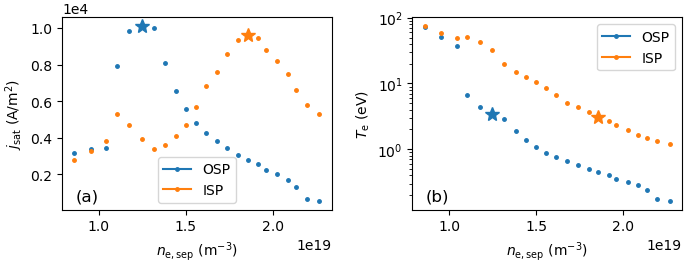}
  \caption{Ion saturation current $j_\mathrm{sat}$ (a) and electron temperature $T_e$ (b) at OSP (blue) and ISP (orange) versus outer midplane separatrix electron density $n_\mathrm{e,sep}$ for cases with $P = 3\,\mathrm{MW}$, $I_p = 700\,\mathrm{kA}$, carbon fraction of $1\%$, and D scaling factor of $1.0$. Stars denote cases at the detachment onset in the outer and inner divertor, respectively.}\label{Fig:jsat_in-out}
\end{figure}

Due to this asymmetry, it is much harder to achieve the roll-over of $j_\mathrm{sat}$ at the inner strike point (inner divertor detachment onset) than the outer divertor, as shown in Fig.~\ref{Fig:onset_inner}.
The inner divertor detachment onset (cases in blue) occurs when the inner strike point temperature $T_\mathrm{e,isp}\sim 3-4\,\mathrm{eV}$, consistent with the outer divertor detachment condition studied above, where $T_\mathrm{e,osp}\sim 3-4\,\mathrm{eV}$ at onset. However, outer divertor detachment onset (green) occurs at much higher $T_\mathrm{e,isp}$. An example of density scan series comparing in-out divertor $j_\mathrm{sat}$ and $T_e$ at $P=3\,\mathrm{MW}$, carbon fraction of $1.3\%$ and D scaling factor of 1.0, is shown in Fig.~\ref{Fig:jsat_in-out}. Outer divertor detachment onset occurs when OSP $j_\mathrm{sat}$ reaches its roll-over point and $T_\mathrm{e,osp}$ drops to $\sim 3\,\mathrm{eV}$ simultaneously, at an outer midplane separatrix density $n_\mathrm{e,sep}\approx 1.25\times 10^{19}\mathrm{m}^{-3}$. In comparison, inner divertor detachment onset requires a higher $n_\mathrm{e,sep}\approx 1.85\times 10^{19}\mathrm{m}^{-3}$, when ISP $j_\mathrm{sat}$ rolls over and $T_\mathrm{e,isp}$ falls similarly to $\sim 3\,\mathrm{eV}$. The similar target $T_e$ values required to reach detachment onset at both the inner and outer targets across a large range of input powers, diffusivities, and carbon fractions, despite strong in-out asymmetry, indicate that target $T_e$ is a robust indicator of detachment, consistent with the findings of Stangeby~\cite{Stangeby2018}.

\begin{figure}
  \centering  \includegraphics[width=1.\textwidth]{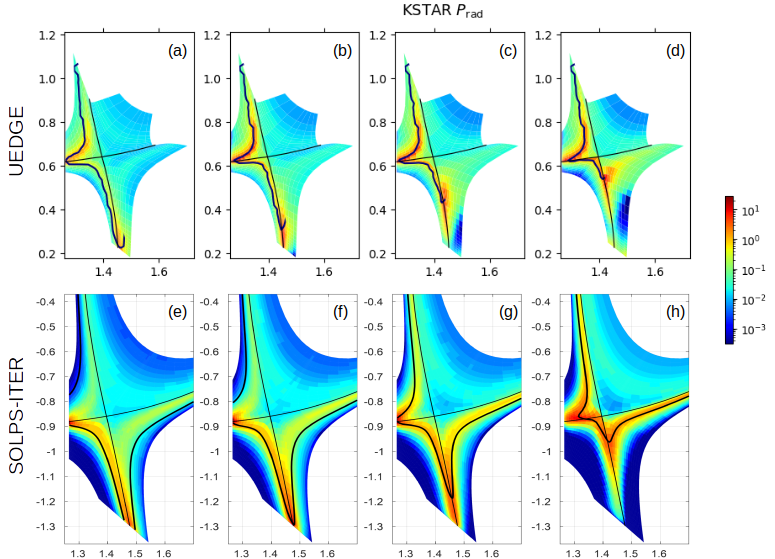}
\caption{Evolution of 2D radiation profile in the divertor region from attached to detached outer divertor both for KSTAR from UEDGE (a,b,c,d) and SOLPS (e,f,g,h) simulations. The upstream density is increased from the left column (a,e) to the right column (d,h) to cover attached (a,e), detach-onset (b,f), deeply detached outer divertor (c and d, g and h). In each panel, contour of $T_e=10\,\mathrm{eV}$ is also shown (black curve).}\label{Fig:Prad_KSTAR}
\end{figure}

\begin{figure}
  \centering 
  \includegraphics[width=1.\textwidth]{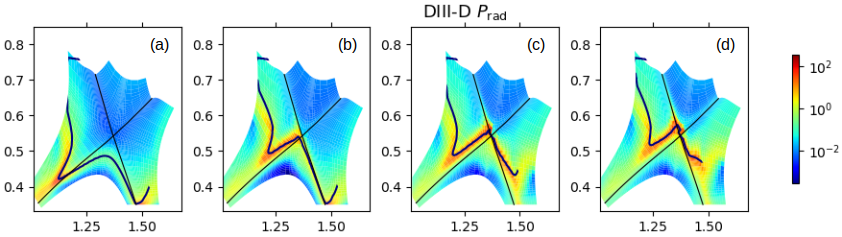}
  \caption{Evolution of 2D radiation profile in the divertor region from attached to detached outer divertor both for DIII-D from UEDGE. In each plot, contour of $T_e=10\,\mathrm{eV}$ is also shown (black curve).}\label{Fig:Prad_DIII-D}
\end{figure}

As shown in Fig.~\ref{Fig:Prad_KSTAR}, the different characteristics of in-out asymmetry in detachment between KSTAR and the other tokamaks like DIII-D, ASDEX-U and JET can be further seen from the evolution of 2D radiation profile from attached to detached in outer divertor.
\begin{itemize}
    \item KSTAR: when the outer divertor is attached $T_\mathrm{e,osp} > 10-30 \,\mathrm{eV}$, the inner and outer divertors are quite symmetric $T_\mathrm{e,osp} \sim T_\mathrm{e,isp}$ (shown as (a) in Fig.~\ref{Fig:Prad_KSTAR}). As upstream density increases, the outer divertor temperature drops much quicker than the inner divertor (shown as (b) in Fig.~\ref{Fig:Prad_KSTAR}). The outer divertor reaches the onset of detachment first at $T_\mathrm{e,osp} \sim 3 \,\mathrm{eV}$ while the inner divertor is still quite attached $T_\mathrm{e,isp} > 10 \,\mathrm{eV}$ (shown as (c) in Fig.~\ref{Fig:Prad_KSTAR}). The LFS radiation front starts moving away from the outer target plate at detachment onset and approaches the X-point when the outer divertor is deeply detached (shown as (d) in Fig.~\ref{Fig:Prad_KSTAR}). This in-out asymmetry in detachment front dynamics seen in UEDGE is consistent with SOLPS-ITER simulations (shown as (e),(f),(g),(h) in Fig.~\ref{Fig:Prad_KSTAR}).
    \item DIII-D: The in-out asymmetry is already present when the outer divertor is attached, i.e. the inner divertor is deeply detached and the HFS radiation front is away from the inner plate while $T_\mathrm{e,osp} \sim 20-30 \,\mathrm{eV}$. As upstream density increases, the HFS radiation front moves towards the X-point along the inner leg while the outer divertor is kept attached. A detachment bifurcation ($T_e$ cliff~\cite{cliff1,cliff2,cliff3,cliff4,cliff5}) exists in the scan series of simulations. It occurs, dropping the temperature in the outer divertor and make it deeply detached, when the HFS radiation front develops inside of the separatrix. The evolution of DIII-D detachment front is shown as (a), (b), (c) and (d) in Fig.~\ref{Fig:Prad_DIII-D}.
\end{itemize}
The unique in–out divertor detachment asymmetry observed in the UEDGE database is consistent with experimental findings reported in~\cite{ParkNF2018}, which attribute the behavior primarily to the KSTAR divertor geometry. Specifically, the vertically oriented inner target with its short poloidal connection from the X-point, together with the long, inclined outer divertor leg, promotes strong neutral accumulation near the outer target. This in turn enhances local momentum and power losses, leading to earlier detachment of the outer divertor compared to the inner.

\subsubsection{Dependence of in-out asymmetry on control parameters.}
\begin{figure}
\centering  \includegraphics[width=.5\textwidth]{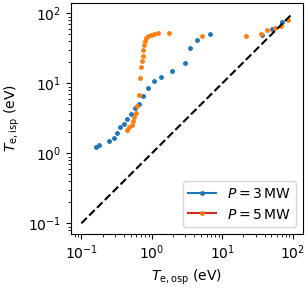}
  \caption{ISP electron temperature $T_\mathrm{e,isp}$ versus OSP electron temperature $T_\mathrm{e,osp}$ for cases with input power at $P = 3\,\mathrm{MW}$ (blue) and $5\,\mathrm{MW}$ (orange), with $I_p = 700\,\mathrm{kA}$, D scaling factor of $1.0$, and carbon fraction of $1.3\%$. The dashed line denotes where $T_\mathrm{e,isp} = T_\mathrm{e,osp}$.}\label{Fig:Tio}
\end{figure}

The general level of in-out divertor asymmetry can be indicated by the ratio of ISP electron temperature to OSP electron temperature $T_\mathrm{e,isp}/T_\mathrm{e,osp}$. This ratio is clearly depending on collisionality, as shown in Fig.~\ref{Fig:Tio}.
When the SOL is collisionless, $T_e$ along a flux tube is nearly isothermal so $T_\mathrm{e,isp}/T_\mathrm{e,osp} \sim 1$. As mentioned above, for $P=3\,\mathrm{MW}$, the outer divertor detaches earlier and $T_\mathrm{e,osp}$ drops to $\sim 3\,\mathrm{eV}$ while the inner divertor is still quite attached with $T_\mathrm{e,isp}\sim 30\,\mathrm{eV}$. This results in $T_\mathrm{e,isp}/T_\mathrm{e,osp} \sim 10$ at onset of outer divertor detachment. With further increase of the upstream density, $T_\mathrm{e,isp}/T_\mathrm{e,osp}$ develops towards $1.0$ once inner and outer divertors achieve detachment.

This characteristics of $T_\mathrm{e,isp}/T_\mathrm{e,osp}$ with upstream density is clearly depending on input power, comparing the scan series with $P = 3\,\mathrm{MW}$ and $P = 5\,\mathrm{MW}$ in Fig.~\ref{Fig:Tio}. The in-out divertor asymmetry is more extreme with $P = 5\,\mathrm{MW}$ at $T_\mathrm{e,osp}\sim 1-5 \mathrm{eV}$, with $T_\mathrm{e,isp}/T_\mathrm{e,osp}$ peaks to $30$.

\begin{figure}
  \centering
  \includegraphics[width=\textwidth]{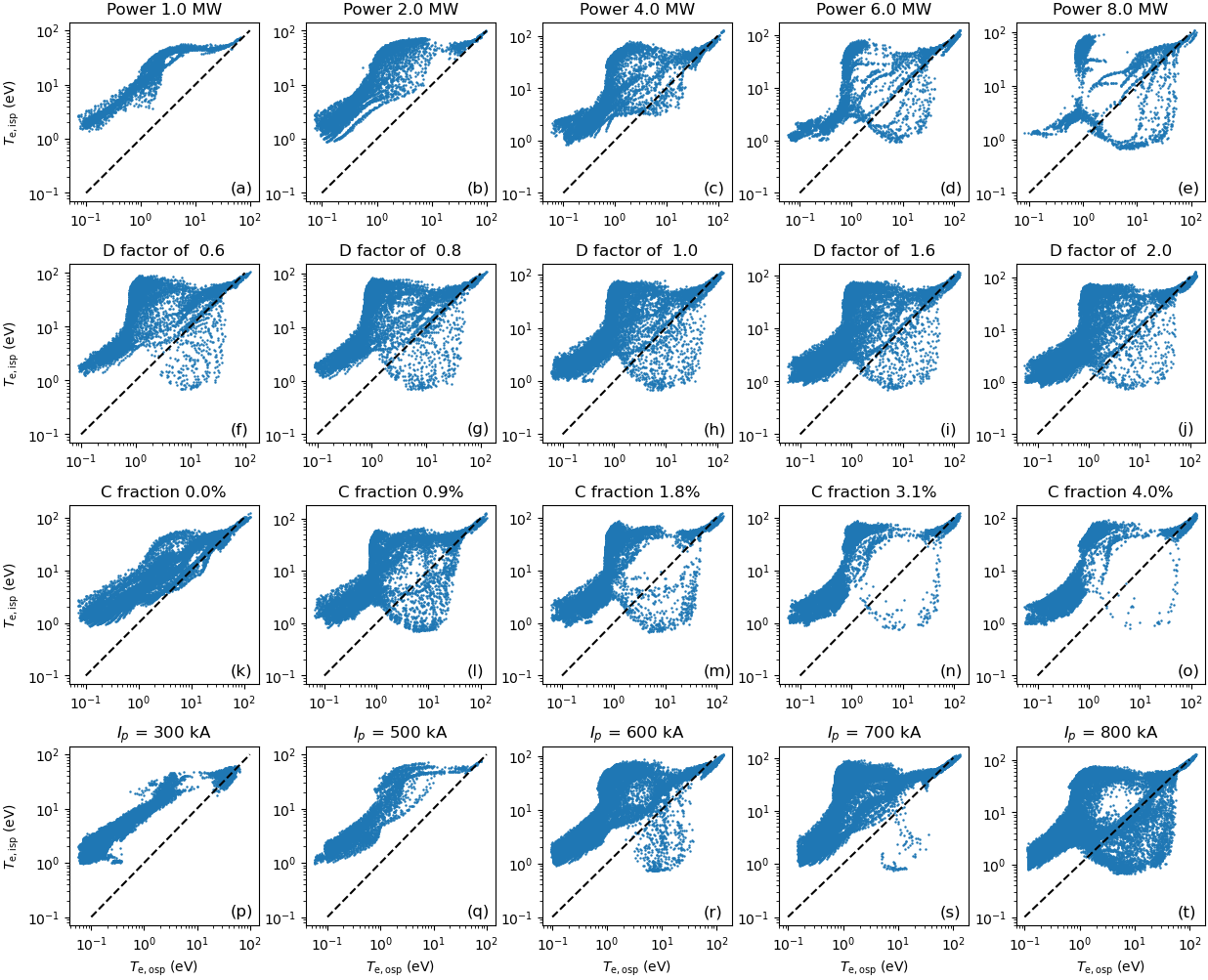}
  \caption{ISP electron temperature $T_\mathrm{e,isp}$ versus OSP electron temperature $T_\mathrm{e,osp}$ for all cases in the database, at input power of $1\,\mathrm{MW}$ (a), $2\,\mathrm{MW}$ (b), $4\,\mathrm{MW}$ (c), $6\,\mathrm{MW}$ (d), $8\,\mathrm{MW}$ (e), with D scaling factor of $0.6$ (f), $0.8$ (g), $1.0$ (h), $1.6$ (i), $2.0$ (j), with carbon fraction of $0\%$ (k), $0.9\%$ (l), $1.8\%$ (m), $3.1\%$ (m). $4.0\%$ (o), with $I_p = 300\,\mathrm{kA}$ (p), $I_p = 500\,\mathrm{kA}$ (q), $I_p = 600\,\mathrm{kA}$ (r), $I_p = 700\,\mathrm{kA}$ (s), $I_p = 800\,\mathrm{kA}$ (t). The dashed line in each plot denotes where $T_\mathrm{e,isp} = T_\mathrm{e,osp}$.}\label{Fig:Tio_dep}
\end{figure}

The dependence of the characteristics of in-out divertor asymmetry ($T_\mathrm{e,isp}/T_\mathrm{e,osp}$) on the control parameters: input power, diffusivity scaling factor, carbon fraction and plasma current is summarized in Fig.~\ref{Fig:Tio_dep}. The results indicate that in–out divertor asymmetry depends strongly on input power and plasma current, but is largely insensitive to diffusivity variation. While there appears to be a dependence on carbon fraction, the converged cases with higher impurity fraction typically correspond to higher power, and those with lower impurity content to lower power, suggesting that impurity fraction effects may instead be linked to power. Therefore, no definitive dependence on impurity fraction can be concluded. We note that while a strong dependence on input power is consistently observed, the underlying physics responsible for this trend remains unresolved. Given the complexity of the problem, a detailed investigation lies beyond the scope of this work but is recognized as an important subject worthy of a dedicated future study.

\begin{figure}
  \centering
  \includegraphics[width=\textwidth]{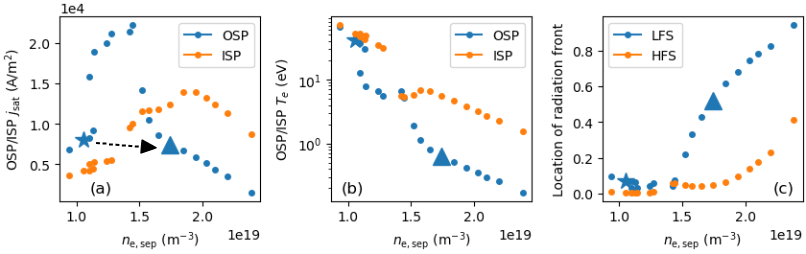}
  \caption{Summary of in–out divertor asymmetry: comparison of OSP (blue) and ISP (orange) ion saturation current, electron temperature, and radiation-front location versus outer midplane separatrix density. The arrows indicate the evolution direction of a density scan. The outer divertor detaches at lower $n_\mathrm{e,sep}$ than the inner divertor, illustrating the earlier onset of OSP detachment.}
\label{Fig:scan_puff}
\end{figure}
Fig.~\ref{Fig:jsat_in-out} and Fig.~\ref{Fig:Prad_KSTAR} demonstrate the distinct onset conditions and radiation profiles of the inner and outer divertors. To complement these results, Fig.~\ref{Fig:scan_puff} provides a consolidated view of OSP and ISP behavior, showing ion saturation current, electron temperature, and the evolution of the radiation front as functions of separatrix density. This summary highlights the earlier onset of detachment in the outer divertor, with $T_\mathrm{e,osp}\sim 3$ eV and $j_\mathrm{sat}$ rollover occurring at lower $n_\mathrm{e,sep}$, while the inner divertor remains attached until significantly higher densities.

\section{Time-Dependent UEDGE Simulations and Intrinsic Detachment Dynamics}
\label{sec:dynamic_database}
The time-dependent UEDGE simulations presented in this section were performed using representative steady-state cases drawn from the parameter scans described in Section \ref{sec:steady_database}, ensuring direct consistency between the dynamic and database-based analyses.

Achieving robust real-time detachment control requires not only an understanding of equilibrium detachment states but also of the transient plasma dynamics that govern the response to actuator variations. While most existing detachment studies focus on steady-state characteristics, recent experiments have demonstrated the critical role of dynamics in shaping achievable control performance~\cite{Timedep1,Timedep2,Timedep3}. In particular, the evolution of radiation fronts and the finite latency of plasma response to gas puffing or impurity seeding must be captured for predictive control algorithms. To this end, complementary time-dependent UEDGE simulations have been performed to quantify the temporal evolution of detachment in KSTAR, with emphasis on the latency and relaxation timescales most relevant for actuator-based control.  

\subsection{Simulation setup}  
The simulations evolve plasma from an attached to a detached state under conditions representative of KSTAR operation ($P_{\mathrm{SOL}}=3$~MW, plasma current $I_p=700$~kA, carbon fraction of 1\%, and transport scaling factor of unity). Instead of prescribing the core boundary density, deuterium fueling is controlled through gas puffing at the outer midplane. Starting from an attached equilibrium with a $D_0$ gas puffing rate of $6.87\times 10^{20}\,\mathrm{s}^{-1}$ (denoted as 'star' in Fig.~\ref{Fig:scan_puff}), the rate is ramped to $2.81\times 10^{21}\,\mathrm{s}^{-1}$ ---corresponding to a deeply detached state (denoted as 'triangle' in Fig.~\ref{Fig:scan_puff})---using four characteristic rise times: $10^{-6}$~ms, $10^{-3}$~ms, $1$~ms, and $100$~ms.  

\subsection{Plasma response to actuator timescales}  
For rapid actuator changes ($\leq 1$~ms), the plasma evolution is largely insensitive to the ramp rate, since the intrinsic plasma response time is $\gtrsim 20\,\mathrm{ms}$. Only when the gas puff is varied on timescales comparable to or slower than this intrinsic response does the actuator dynamics become evident in the plasma trajectories. This highlights a fundamental latency barrier: variations faster than a few milliseconds are effectively invisible to the plasma, while slower ramps reveal the characteristic relaxation processes. The collapse of trajectories for rise times $\lesssim 1$\,ms and the actuator-limited behavior for the slowest ramp are evident in Fig.~\ref{fig:rise-times-multicharge}.

The temperature at the outer strike point drops from $\sim 30$~eV to $\sim 8$~eV on a timescale of roughly 10 ms (ranging from about 5 to 15 ms across cases), nearly independent of actuator variation, suggesting the presence of an intrinsic transient plasma dynamics rather than by external control.

\subsection{Impurity model validation}  
\begin{figure}
  \includegraphics[width=.47\textwidth]{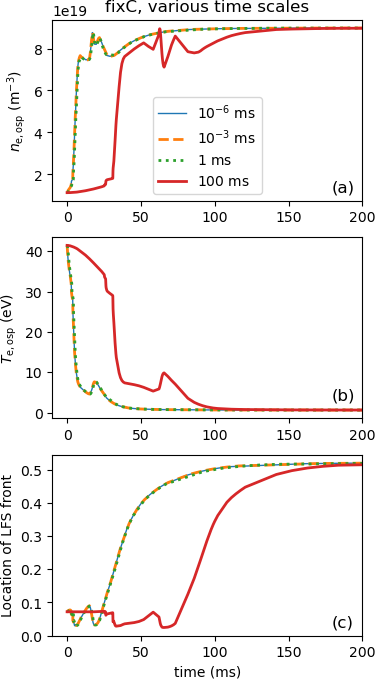}
  \hfill
  \includegraphics[width=.48\textwidth]{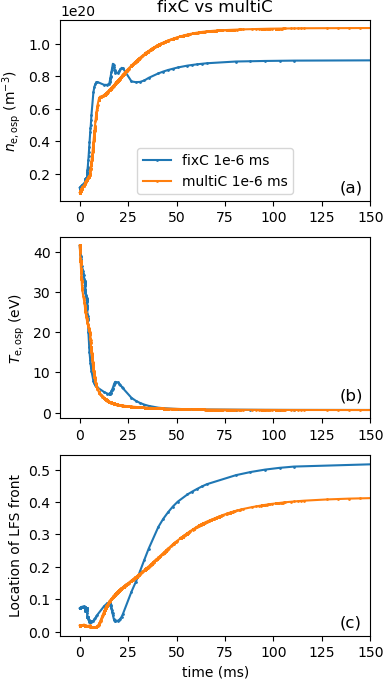}
  \caption{(Left) Actuator-timescale scan: plasma response to outer–midplane gas-puff ramps from $6.87\times 10^{20}\,\mathrm{s}^{-1}$ to $2.81\times 10^{21}\,\mathrm{s}^{-1}$ with rise times $10^{-6}$–100\,ms at $P_{\mathrm{SOL}}=3$\,MW, $I_p=700$\,kA, $f_{\mathrm{C}}=1\%$, and $D$-scale$=1$. For rise times $\lesssim1$\,ms, trajectories are indistinguishable, consistent with an intrinsic response $\gtrsim20$\,ms; only the 100\,ms ramp reveals actuator-limited behavior. (Right) Comparison of fixed-fraction (blue) and multi–charge-state (red) carbon models: the latter yields smoother traces but consistent FOPDT parameters $(L,\tau)$, confirming that the fixed-fraction model captures the essential detachment dynamics.}\label{fig:rise-times-multicharge}
\end{figure}

To verify that the simplified fixed-fraction carbon model does not introduce artifacts in the dynamic response, additional simulations were performed using a multi–charge-state impurity model. The latter yields smoother temporal trajectories, without the kink-like features seen in the fixed-fraction case, but produces nearly identical delay times and relaxation constants. The close agreement of the extracted FOPDT parameters $(L,\tau)$ between the two impurity treatments [Fig.~\ref{fig:rise-times-multicharge}(right)] confirms that detachment dynamics are robust to the impurity model choice, validating the fixed-fraction approximation for large-scale database generation.

\subsection{Reduced-order representation with FOPDT models}  
\label{sec:time_domain_fit}
To enable compact, control-relevant descriptions of plasma dynamics, the transient trajectories of electron density and temperature at four key locations—outer strike point (OSP), inner strike point (ISP), outer midplane separatrix, and outer midplane core boundary together with the position of the low-field-side (LFS) radiation front, were fitted using a first-order-plus-dead-time (FOPDT) model. For a step input applied at $t=0$, the system response is approximated by  
\begin{equation}
\label{eq:step-response}
y(t) = y_\infty - \left(y_\infty - y_0\right) 
\exp\!\left(-\frac{t-L}{\tau}\right), \quad t>L,
\end{equation}
where $L$ is the input delay (dead time), $\tau$ is the response time, and $y_\infty$ is the steady-state value. 
\begin{figure}
  \centering
  \includegraphics[width=.45\textwidth]{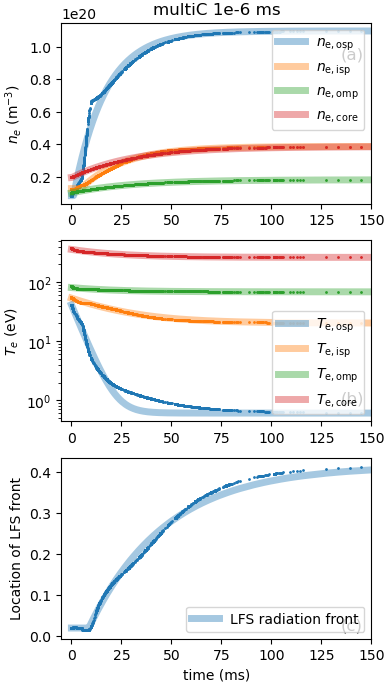}
  \caption{Electron density (a) and temperature (b) at OSP (\textcolor{C0}{blue}), ISP (\textcolor{C1}{orange}), OMP Separatrix (\textcolor{C2}{green}), OMP core boundary (\textcolor{C3}{red}), and location of the LFS radiation front (c) as functions of time with $D_0$ gas puff rate increased from $6.87\times 10^{20}\,\mathrm{s}^{-1}$ to $2.81\times 10^{21}\,\mathrm{s}^{-1}$ in a time scale of $10^{-6}\,\mathrm{ms}$ with the multi-charged C model. Dots: time-dependent UEDGE simulations with adaptive time steps. Curves: FOPDT fitting of the UEDGE time-dependent data.}\label{Fig:time_fit}
\end{figure}
This compact form represents the standard first-order-plus-dead-time (FOPDT) model widely used in process control to approximate delayed first-order responses~\cite{NormeyRico2007, Seborg2016}. Here, $(L,\tau)$ are determined by fitting the time evolution of $T_e$, $n_e$, and the radiation-front position following a step-like increase in the gas puff, as shown in Fig.~\ref{Fig:time_fit}. 
The extracted FOPDT parameters $(L,\tau)$ provide a compact description of the intrinsic plasma response, see Fig.~\ref{fig:charactertime}, which can be directly incorporated into control design and frequency-domain analysis, as discussed in Section~4.5. Representative intrinsic values summarized in Table~\ref{tab:FOPDT-summary} highlight the spatial ordering of response times—fastest at the OSP and slowest for the LFS radiation front—and provide the quantitative basis for control analysis discussed next in Section~\ref{sec:control_design}.
\begin{figure}
  \includegraphics[width=\textwidth]{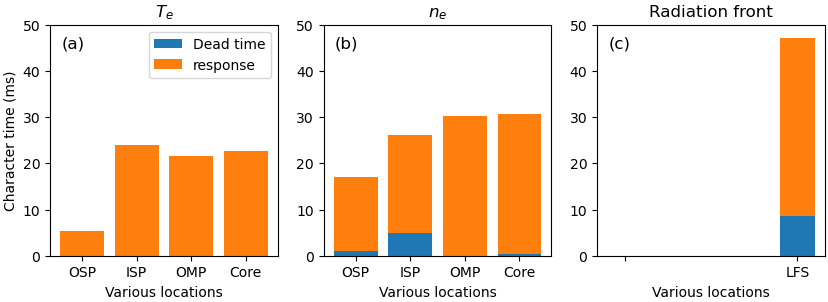}
\caption{
FOPDT fits~(15) to (a) local divertor $T_e$, (b) local divertor $n_e$, and (c) the low-field-side radiation-front position (defined by the peak location of the radiation) following a gas-puff step. Bars indicate fitted timescales: \textcolor{blue}{blue} = dead time $L$, \textcolor{orange}{orange} = response time $\tau$. The radiation front in (c) exhibits the longest latency—$L\!\sim\!10$ ms and $\tau\!\sim\!40$ ms—substantially slower than the local divertor responses in (a,b). The extracted $(L,\tau)$ values represent the \emph{intrinsic plasma response} obtained from UEDGE simulations and provide the basis for reduced dynamic models used in control-system analysis (see Section~\ref{sec:control_design}). For comparison, experimentally derived FOPDT parameters from closed-loop KSTAR detachment-control experiments~\cite{GuptaArXiv2025} include actuator and diagnostic delays, yielding much longer apparent timescales.}
\label{fig:charactertime}
\end{figure}
\begin{table}[htbp]
\centering
\caption{Representative intrinsic FOPDT parameters extracted from UEDGE step-response fits for different plasma locations. 
Listed values correspond to typical cases at $P_{\mathrm{SOL}}\!=\!3$~MW and $I_p\!=\!0.7$~MA. All quantities represent \emph{intrinsic plasma} delays and response times from the UEDGE simulations, excluding actuator and diagnostic latencies. These intrinsic plasma delays and response times are quantified using the averaged response of the local electron temperature and density. Quoted $\pm$ values represent typical variability ($\sim20–30 \%$) among multiple UEDGE time-dependent runs and fit residuals, rather than formal statistical errors.}
\vspace{4pt}
\begin{tabular}{lcc}
\hline
\textbf{Location} & \textbf{Dead time $L$ (ms)} & \textbf{Response time $\tau$ (ms)} \\
\hline
Outer strike point (OSP)           & $2 \pm 2$     & $15 \pm 10$ \\
Inner strike point (ISP)           & $5 \pm 5$     & $25 \pm 5$ \\
Outer midplane separatrix (OMP-sep) & $1 \pm 1$     & $30 \pm 10$ \\
LFS radiation front (10 eV contour) & $10 \pm 3$    & $40 \pm 5$ \\
\hline
\end{tabular}
\label{tab:FOPDT-summary}
\end{table}

\subsection{Implications for control design}
\label{sec:control_design}
Building on the intrinsic fits summarized in Table~\ref{tab:FOPDT-summary}, the FOPDT parameters $(L,\tau)$ extracted in Section~\ref{sec:time_domain_fit} provide a compact dynamic description of the plasma response and form the basis 
for the control-system representations discussed below.

In feedback control, this behavior is commonly represented in the \emph{Laplace domain} by the standard first-order-plus-dead-time (FOPDT) transfer function,
\begin{equation}
\label{eq:laplace_fopdt}
G(s) = \frac{K\,e^{-L s}}{1+\tau s},
\end{equation}
where $s=\sigma+j\omega$ is the complex Laplace variable and $K=\partial y/\partial u$ denotes the local steady-state (small-signal) gain between the actuator input~$u$ (e.g.\ gas-puff command) and the controlled variable~$y$ (e.g.\ $T_e$, $n_e$, or radiation-front position). The exponential term $e^{-Ls}$ represents a pure time delay, while $(1+\tau s)^{-1}$ captures the first-order relaxation dynamics.

Evaluating Eq.~\eqref{eq:laplace_fopdt} on the imaginary axis ($s=j\omega$) yields the \emph{frequency-domain} or Bode representation,
\begin{equation}
\label{eq:freq_fopdt}
|G(j\omega)| = \frac{K}{\sqrt{1+(\omega\tau)^2}}, \qquad
\angle G(j\omega) = -\!\left[\omega L + \tan^{-1}(\omega\tau)\right],
\end{equation}
where the negative sign denotes phase lag according to the standard control-engineering convention also used by Gupta~\textit{et al.}~\cite{GuptaArXiv2025}. The total phase decreases with increasing frequency because of both the intrinsic relaxation ($\tau$) and the finite delay ($L$), thereby limiting achievable control bandwidth and gain margin.

In the KSTAR detachment-control experiments of Gupta \textit{et al.}~\cite{GuptaArXiv2025}, similar FOPDT models were fitted to frequency-response measurements between the gas-valve command and diagnostic signals. Two representative operating modes were characterized: (i) for the attachment-fraction ($A_{\mathrm{frac}}$) controller, $L_{\mathrm{exp}}\!\approx\!0.15$–0.20 s and $\tau_{\mathrm{exp}}\!\approx\!0.4$ s; and (ii) for the surrogate-model (\textit{DivControlNN}) controller, $L_{\mathrm{exp}}\!\approx\!0.53$ s and $\tau_{\mathrm{exp}}\!\approx\!1.2$ s. These experimental values were obtained by applying variations to the gas-puff rate that included step-like increases and decreases, as shown in Fig.~5 of Gupta~\textit{et al.}~\cite{GuptaArXiv2025}. The resulting plasma response was then fitted to Eq.~(17) using least-squares regression to extract the FOPDT parameters.

The intrinsic parameters extracted from the present UEDGE simulations ($L_{\mathrm{sim}}\!\sim\!5$--10~ms, $\tau_{\mathrm{sim}}\!\sim\!5$--40~ms) are more than an order of magnitude shorter than the experimental values. This difference is expected: the simulations isolate the \emph{plasma-only} dynamics, whereas the experimentally measured response incorporates additional latencies from the actuator, gas-line transport, diagnostic integration, and digital-controller update.  Thus,
\begin{equation}
\label{eq:L-comp-exp-plasma}
L_{\mathrm{exp}} \;=\;
L_{\mathrm{plasma}} + L_{\mathrm{act}} + L_{\mathrm{diag}} + L_{\mathrm{ctrl}},
\end{equation}
with the present UEDGE results representing the intrinsic $L_{\mathrm{plasma}}$ component. Although the absolute timescales differ, the dimensionless ratio $L/\tau$---typically $0.2$--$0.5$ in both simulation and experiment---is comparable, indicating a consistent dynamical regime in which delay effects dominate over inertial relaxation. This agreement suggests that both the simulated and measured systems are \emph{delay-limited} rather than inertia-limited.

The fitted FOPDT parameters $(K,L,\tau)$ therefore supply a unified physics-based foundation for controller design.  They can be used to determine proportional-integral (PI) gains through standard tuning rules, to define prediction horizons in model-predictive control (MPC), and to estimate achievable control bandwidths. Because $K$ and $\tau$ vary with the detachment state, gain-scheduled or state-dependent controllers—as implemented by Gupta~\textit{et al.}—are appropriate for robust operation across attached, transition, and detached regimes.

\medskip  
\noindent\textbf{Summary of Section 4:} This section characterized intrinsic detachment dynamics from time-dependent UEDGE simulations using first-order-plus-dead-time (FOPDT) models. The fitted parameters $(L,\tau)$ exhibit clear spatial variation—from a few milliseconds at the OSP to tens of milliseconds for the LFS radiation front—quantifying the plasma’s intrinsic latency and relaxation behavior. These results establish a direct link between detailed UEDGE simulations and reduced dynamic models for real-time control. The extracted $(L,\tau)$ values provide a quantitative basis for extending existing surrogate controllers such as DivControlNN~\cite{Zhu} and for refining latency-compensation and tuning strategies in detachment-control experiments such as those of Gupta~\textit{et~al.}~\cite{GuptaArXiv2025}, enabling next-generation predictive control in KSTAR and related devices.

\section{summary}
A comprehensive UEDGE database has been generated to support surrogate-model development and improve detachment-control algorithms in KSTAR. Nearly 70{,}000 steady-state solutions were produced by scanning core boundary density, input power, impurity fraction, diffusivity scaling, and plasma current across the KSTAR operational space, with cross-field drifts included for fidelity. From this database, several robust detachment characteristics emerge. The strike-point electron temperature consistently falls near 3--4~eV at detachment onset, largely independent of upstream density, input power, impurity fraction, diffusivity, or plasma current. At onset, the low-field-side radiation front shifts toward the strike point and then moves poloidally away, providing a clear diagnostic signature.

Scaling laws derived from the 2D database show weaker impurity sensitivity than 1D models and reveal an explicit dependence on cross-field transport, with heat-flux widths ($\lambda_q$) following Eich’s scaling only for uniform, low $D/\chi$. KSTAR exhibits distinctive in--out divertor asymmetries: the outer divertor detaches earlier and more deeply than the inner, opposite to trends seen in DIII-D, ASDEX-U, and JET. This difference is linked to KSTAR’s divertor geometry and is consistent with prior SOLPS-ITER modeling and experiment.

Complementary time-dependent simulations quantify plasma response to gas puffing, highlighting finite delays that shape achievable control bandwidths. Response times are $\sim$5--15~ms at the outer strike point, $\sim$25--30~ms at the midplane and inner divertor, and $\sim$40~ms for the low-field-side radiation front, with an additional $\sim$10~ms input delay. These dynamics are well captured by first-order-plus-dead-time (FOPDT) models, enabling reduced representations that preserve both equilibrium mappings and latency effects.

Together, the steady-state database and dynamic simulations establish a physics-based foundation for predictive detachment control in KSTAR. They provide equilibrium-based mappings, scaling laws, and compact dynamic models that directly support machine-learning surrogate development and real-time control strategies. Future work will extend the database to multi-charge-state impurity models, improve low-current coverage, and validate against upcoming KSTAR detachment experiments.

The FOPDT-based dynamic characterization developed here has been experimentally validated in the KSTAR detachment-control experiments~\cite{GuptaArXiv2025}, confirming that the intrinsic plasma response times (5--40~ms) identified from UEDGE simulations underlie the slower, system-level dynamics observed in real-time control.

\section{Acknowledgments}
This work was performed under the U.S. Department of Energy by Lawrence Livermore National Laboratory under Contract No. DE-AC52-07NA27344, LLNL-JRNL-2011854. The authors gratefully acknowledge valuable discussions with Timo Bremer, H. Bhatia, and the UEDGE team.

\section{Reference}

\bibliographystyle{unsrt} 
\bibliography{uedge_database}

\begin{thebibliography}{10}

\bibitem{10MW}
R.A. Pitts, X.~Bonnin, F.~Escourbiac, H.~Frerichs, J.P. Gunn, T.~Hirai, A.S. Kukushkin, E.~Kaveeva, M.A. Miller, D.~Moulton, V.~Rozhansky, I.~Senichenkov, E.~Sytova, O.~Schmitz, P.C. Stangeby, G.~{De Temmerman}, I.~Veselova, and S.~Wiesen.
\newblock Physics basis for the first iter tungsten divertor.
\newblock {\em Nuclear Materials and Energy}, 20:100696, 2019.

\bibitem{detachment1}
A~W Leonard.
\newblock Plasma detachment in divertor tokamaks.
\newblock {\em Plasma Physics and Controlled Fusion}, 60(4):044001, feb 2018.

\bibitem{detachment2}
S.~I. Krasheninnikov, A.~S. Kukushkin, and A.~A. Pshenov.
\newblock Divertor plasma detachment.
\newblock {\em Physics of Plasmas}, 23(5):055602, 05 2016.

\bibitem{TCV}
Perek~A. Ravensbergen~T., van Berkel~M. et~al.
\newblock Real-time feedback control of the impurity emission front in tokamak divertor plasmas.
\newblock {\em Nat Commun}, 12:1105, 2021.

\bibitem{DIII-D}
D.~Eldon, E.~Kolemen, J.L. Barton, A.R. Briesemeister, D.A. Humphreys, A.W. Leonard, R.~Maingi, M.A. Makowski, A.G. McLean, A.L. Moser, and P.C. Stangeby.
\newblock Controlling marginally detached divertor plasmas.
\newblock {\em Nuclear Fusion}, 57(6):066039, may 2017.

\bibitem{KSTAR}
D~Eldon, H~Anand, J-G Bak, J~Barr, S-H Hahn, J~H Jeong, H-S Kim, H~H Lee, A~W Leonard, B~Sammuli, G~W Shin, and H~Q Wang.
\newblock Enhancement of detachment control with simplified real-time modelling on the kstar tokamak.
\newblock {\em Plasma Phys. Control. Fusion}, 64(7):075002, may 2022.

\bibitem{EAST}
D.~Eldon, H.Q. Wang, L.~Wang, J.~Barr, S.~Ding, A.~Garofalo, X.Z. Gong, H.Y. Guo, A.E. Järvinen, K.D. Li, J.~McClenaghan, A.G. McLean, C.M. Samuell, J.G. Watkins, D.~Weisberg, and Q.P. Yuan.
\newblock An analysis of controlled detachment by seeding various impurity species in high performance scenarios on diii-d and east.
\newblock {\em Nuclear Materials and Energy}, 27:100963, 2021.

\bibitem{MANTIS}
A.~Perek, W.~A.~J. Vijvers, Y.~Andrebe, I.~G.~J. Classen, B.~P. Duval, C.~Galperti, J.~R. Harrison, B.~L. Linehan, T.~Ravensbergen, K.~Verhaegh, M.~R. de~Baar, TCV, and EUROfusion~MST1 Teams.
\newblock Mantis: A real-time quantitative multispectral imaging system for fusion plasmas.
\newblock {\em Review of Scientific Instruments}, 90(12):123514, 12 2019.

\bibitem{Zhu2}
Ben Zhu, Menglong Zhao, Harsh Bhatia, Xue-qiao Xu, Peer-Timo Bremer, William Meyer, Nami Li, and Thomas Rognlien.
\newblock Data-driven model for divertor plasma detachment prediction.
\newblock {\em Journal of Plasma Physics}, 88(5):895880504, 2022.

\bibitem{Zhu}
Ben Zhu, Menglong Zhao, Xue-Qiao Xu, Anchal Gupta, KyuBeen Kwon, Xinxing Ma, and David Eldon.
\newblock Latent space mapping: Revolutionizing predictive models for divertor plasma detachment control.
\newblock {\em Physics of Plasmas}, 32(6):062508, 06 2025.

\bibitem{GuptaArXiv2025}
Anchal Gupta, David Eldon, Eunnam Bang, KyuBeen Kwon, Hyungho Lee, Anthony Leonard, Junghoo Hwang, Xueqiao Xu, Menglong Zhao, and Ben Zhu.
\newblock Detachment control in kstar with tungsten divertor, 2025.

\bibitem{Stangeby2018}
PC~Stangeby.
\newblock Basic physical processes and reduced models for plasma detachment.
\newblock {\em Plasma Physics and Controlled Fusion}, 60(4):044022, 2018.

\bibitem{Goldston}
R~J Goldston, M~L Reinke, and J~A Schwartz.
\newblock A new scaling for divertor detachment.
\newblock {\em Plasma Physics and Controlled Fusion}, 59(5):055015, mar 2017.

\bibitem{Body}
Thomas Body, Thomas Eich, Adam Kuang, Tom Looby, Mike Kryjak, Ben Dudson, and Matthew Reinke.
\newblock Detachment scalings derived from 1d scrape-off-layer simulations.
\newblock {\em Nuclear Materials and Energy}, 41:101819, 2024.

\bibitem{Eich}
T.~Eich, A.W. Leonard, R.A. Pitts, W.~Fundamenski, R.J. Goldston, T.K. Gray, A.~Herrmann, A.~Kirk, A.~Kallenbach, O.~Kardaun, A.S. Kukushkin, B.~LaBombard, R.~Maingi, M.A. Makowski, A.~Scarabosio, B.~Sieglin, J.~Terry, A.~Thornton, ASDEX~Upgrade Team, and JET~EFDA Contributors.
\newblock Scaling of the tokamak near the scrape-off layer h-mode power width and implications for iter.
\newblock {\em Nuclear Fusion}, 53(9):093031, aug 2013.

\bibitem{XuPSI2025}
XQ~Xu, NM~Li, ML~Zhao, X~Liu, PH~Diamond, B~Zhu, TD~Rognlien, and GS~Xu.
\newblock Fluctuation entrainment and sol width broadening in small/grassy elm regime.
\newblock {\em Nuclear Materials and Energy}, 42:101866, 2025.

\bibitem{XuNF2019}
XQ~Xu, NM~Li, ZY~Li, B~Chen, TY~Xia, TF~Tang, B~Zhu, and VS~Chan.
\newblock Simulations of tokamak boundary plasma turbulence transport in setting the divertor heat flux width.
\newblock {\em Nuclear Fusion}, 59(12):126039, 2019.

\bibitem{Henderson2019NME}
S.~S. Henderson, M.~Wischmeier, R.~A. Pitts, X.~Bonnin, H.~Frerichs, A.~Kukushkin, D.~Moulton, S.~Wiesen, B.~Lipschultz, and A.~Loarte.
\newblock Comparison of detachment behaviour in jet-ilw and aug with implications for iter.
\newblock {\em Nucl. Mater. Energy}, 18:147--152, 2019.

\bibitem{Henderson2023NF}
S.~S. Henderson, M.~Wischmeier, A.~Loarte, S.~Potzel, H.~Frerichs, A.~Kukushkin, R.~A. Pitts, E.~Wolfrum, and S.~Wiesen.
\newblock Detachment and re-attachment dynamics with mixed impurity seeding in asdex upgrade.
\newblock {\em Nucl. Fusion}, 63:086024, 2023.

\bibitem{Wang2023NF}
W.~Wang, N.~H. Brooks, D.~Eldon, M.~E. Fenstermacher, P.~C. Stangeby, X.~Ma, R.~Maingi, T.~Petrie, and A.~Loarte.
\newblock Experimental and modelling studies of detachment control in diii-d closed divertor configurations.
\newblock {\em Nucl. Fusion}, 63:046004, 2023.

\bibitem{Reinke2017NF}
M.~L. Reinke.
\newblock Power exhaust scaling for tokamak divertors with emphasis on high-field side and device-size effects.
\newblock {\em Nucl. Fusion}, 57:034004, 2017.

\bibitem{Pitts2013JNM}
R.~A. Pitts, S.~Carpentier, F.~Escourbiac, T.~Hirai, V.~Komarov, A.~Kukushkin, A.~Loarte, M.~Merola, A.~S. Naik, G.~Pintsuk, V.~Podkovyrov, M.~Sugihara, B.~Bazylev, and P.~C. Stangeby.
\newblock A full tungsten divertor for iter: physics issues and design status.
\newblock {\em J. Nucl. Mater.}, 438:S48--S56, 2013.

\bibitem{Stangeby2022NF}
P.~C. Stangeby, J.~D. Lore, R.~A. Pitts, J.~M. Canik, and X.~Bonnin.
\newblock A reduced model for the iter divertor based on solps solutions for iter q = 10 baseline conditions.
\newblock {\em Nucl. Fusion}, 62:126058, 2022.

\bibitem{Pitts2019NME}
R.~A. Pitts, X.~Bonnin, F.~Escourbiac, H.~Frerichs, J.~P. Gunn, T.~Hirai, A.~S. Kukushkin, E.~Kaveeva, M.~A. Miller, D.~Moulton, V.~Rozhansky, I.~Senichenkov, E.~Sytova, O.~Schmitz, P.~C. Stangeby, G.~De Temmerman, I.~Veselova, and S.~Wiesen.
\newblock Physics basis for the first iter tungsten divertor.
\newblock {\em Nucl. Mater. Energy}, 20:100696, 2019.

\bibitem{Henderson2022NME}
S.S. Henderson, M.~Bernert, C.~Giroud, D.~Brida, M.~Cavedon, P.~David, R.~Dux, J.R. Harrison, A.~Huber, A.~Kallenbach, J.~Karhunen, B.~Lomanowski, G.~Matthews, A.~Meigs, R.A. Pitts, F.~Reimold, M.L. Reinke, S.~Silburn, N.~Vianello, S.~Wiesen, and M.~Wischmeier.
\newblock Parameter dependencies of the experimental nitrogen concentration required for detachment on asdex upgrade and jet.
\newblock {\em Nuclear Materials and Energy}, 28:101000, 2021.

\bibitem{Ding2024Nature}
S.~Ding, A.~M. Garofalo, H.~Q. Wang, D.~B. Weisberg, Z.~Y. Li, X.~Jian, D.~Eldon, B.~S. Victor, A.~Marinoni, Q.~M. Hu, I.~S. Carvalho, T.~Odstr{\v{c}}il, L.~Wang, A.~W. Hyatt, T.~H. Osborne, X.~Z. Gong, J.~P. Qian, J.~Huang, J.~McClenaghan, C.~T. Holcomb, and J.~M. Hanson.
\newblock A high-density and high-confinement tokamak plasma regime for fusion energy.
\newblock {\em Nature}, 629:555--562, 2024.

\bibitem{Thome2024}
K~E Thome, M~E Austin, A~Hyatt, A~Marinoni, A~O Nelson, C~Paz-Soldan, F~Scotti, W~Boyes, L~Casali, C~Chrystal, S~Ding, X~D Du, D~Eldon, D~Ernst, R~Hong, G~R McKee, S~Mordijck, O~Sauter, L~Schmitz, J~L Barr, M~G Burke, S~Coda, T~B Cote, M~E Fenstermacher, A~Garofalo, F~O Khabanov, G~J Kramer, C~J Lasnier, N~C Logan, P~Lunia, A~G McLean, M~Okabayashi, D~Shiraki, S~Stewart, Y~Takemura, D~D Truong, T~Osborne, M~A Van~Zeeland, B~S Victor, H~Q Wang, J~G Watkins, W~P Wehner, A~S Welander, T~M Wilks, J~Yang, G~Yu, L~Zeng, and the DIII-D~Team.
\newblock Overview of results from the 2023 diii-d negative triangularity campaign.
\newblock {\em Plasma Physics and Controlled Fusion}, 66(10):105018, sep 2024.

\bibitem{ZLi2019}
Ze-Yu Li, X.Q. Xu, Na-Mi Li, V.S. Chan, and Xiao-Gang Wang.
\newblock Prediction of divertor heat flux width for iter using bout++ transport and turbulence module.
\newblock {\em Nuclear Fusion}, 59(4):046014, feb 2019.

\bibitem{Nami2020}
N.~M. Li, X.~Q. Xu, J.~W. Hughes, J.~L. Terry, J.~Z. Sun, and D.~Z. Wang.
\newblock Simulations of divertor heat flux width using transport code with cross-field drifts under the bout++ framework.
\newblock {\em AIP Advances}, 10(1):015222, 01 2020.

\bibitem{Chang2017}
C.S. Chang, S.~Ku, A.~Loarte, V.~Parail, F.~Köchl, M.~Romanelli, R.~Maingi, J.-W. Ahn, T.~Gray, J.~Hughes, B.~LaBombard, T.~Leonard, M.~Makowski, and J.~Terry.
\newblock Gyrokinetic projection of the divertor heat-flux width from present tokamaks to iter.
\newblock {\em Nuclear Fusion}, 57(11):116023, aug 2017.

\bibitem{HFHD}
F.~Reimold, M.~Wischmeier, S.~Potzel, L.~Guimarais, D.~Reiter, M.~Bernert, M.~Dunne, and T.~Lunt.
\newblock The high field side high density region in solps-modeling of nitrogen-seeded h-modes in asdex upgrade.
\newblock {\em Nuclear Materials and Energy}, 12:193--199, 2017.
\newblock Proceedings of the 22nd International Conference on Plasma Surface Interactions 2016, 22nd PSI.

\bibitem{JET2015}
AV~Chankin, G~Corrigan, Mathias Groth, PC~Stangeby, et~al.
\newblock Influence of the e$\times$ b drift in high recycling divertors on target asymmetries.
\newblock {\em Plasma Physics and Controlled Fusion}, 57(9):095002, 2015.

\bibitem{JarvinenNME2017}
AE~Jaervinen, SL~Allen, M~Groth, AG~McLean, TD~Rognlien, CM~Samuell, A~Briesemeister, M~Fenstermacher, DN~Hill, AW~Leonard, et~al.
\newblock Interpretations of the impact of cross-field drifts on divertor flows in diii-d with uedge.
\newblock {\em Nuclear Materials and Energy}, 12:1136--1140, 2017.

\bibitem{MaNF2021}
Xinxing Ma, Huiqian~Q Wang, Houyang~Y Guo, Peter~C Stangeby, ET~Meier, Morgan~W Shafer, and Dan~M Thomas.
\newblock First evidence of dominant influence of e$\times$ b drifts on plasma cooling in an advanced slot divertor for tokamak power exhaust.
\newblock {\em Nuclear Fusion}, 61(5):054002, 2021.

\bibitem{MaurizioNF2024}
R~Maurizio, D~Thomas, JH~Yu, T~Abrams, AW~Hyatt, J~Herfindal, A~Leonard, X~Ma, AG~McLean, J~Ren, et~al.
\newblock Experiments on plasma detachment in a v-shaped slot divertor in the diii-d tokamak.
\newblock {\em Nuclear Fusion}, 64(8):086048, 2024.

\bibitem{ScottiNT}
Filippo Scotti, Alessandro Marinoni, Adam~G McLean, Andrew~Oakleigh Nelson, Carlos Paz-Soldan, Kathreen~E Thome, Menglong Zhao, Steve~L Allen, Max~E Austin, M~Galen Burke, et~al.
\newblock Divertor characterization and access to dissipative divertor conditions in negative triangularity discharges in diii-d.
\newblock {\em Plasma Physics and Controlled Fusion}, 2025.

\bibitem{cliff1}
AE~Jaervinen, SL~Allen, D~Eldon, ME~Fenstermacher, Mathias Groth, David~N Hill, AW~Leonard, AG~McLean, GD~Porter, TD~Rognlien, et~al.
\newblock E$\times$ b flux driven detachment bifurcation in the diii-d tokamak.
\newblock {\em Physical review letters}, 121(7):075001, 2018.

\bibitem{cliff2}
AG~McLean, AW~Leonard, MA~Makowski, M~Groth, SL~Allen, JA~Boedo, BD~Bray, AR~Briesemeister, TN~Carlstrom, D~Eldon, et~al.
\newblock Electron pressure balance in the sol through the transition to detachment.
\newblock {\em Journal of Nuclear Materials}, 463:533--536, 2015.

\bibitem{cliff3}
Hailong Du, Guoyao Zheng, Houyang Guo, Aaro~E Jaervinen, Xuru Duan, Xavier Bonnin, David Eldon, and Dezhen Wang.
\newblock Solps analysis of the necessary conditions for detachment cliff.
\newblock {\em Nuclear Fusion}, 60(4):046028, 2020.

\bibitem{cliff4}
Xinxing Ma, HQ~Wang, HY~Guo, A~Leonard, Roberto Maurizio, ET~Meier, Jun Ren, PC~Stangeby, Greg Sinclair, Daniel~M Thomas, et~al.
\newblock E$\times$ b flow driven electron temperature bifurcation in a closed slot divertor with ion b$\times\nabla$ b away from the x-point in the diii-d tokamak.
\newblock {\em Nuclear Fusion}, 62(12):126048, 2022.

\bibitem{cliff5}
M~Zhao, F~Scotti, TD~Rognlien, AG~McLean, G~Burke, and A~Holm.
\newblock 2d analysis of tokamak divertor-plasma detachment-bifurcation with operational parameters and geometries.
\newblock {\em Nuclear Materials and Energy}, 41:101811, 2024.

\bibitem{ParkNF2018}
Jae~Sun Park, Mathias Groth, Richard Pitts, Jun-Gyo Bak, S.G. Thatipamula, June-Woo Juhn, Suk-Ho Hong, and Wonho Choe.
\newblock Atomic processes leading to asymmetric divertor detachment in kstar l-mode plasmas.
\newblock {\em Nuclear Fusion}, 58(12):126033, nov 2018.

\bibitem{Timedep1}
Timo Ravensbergen, Matthijs van Berkel, Artur Perek, C~Galperti, BP~Duval, O~F{\'e}vrier, RJR Van~Kampen, F~Felici, JT~Lammers, Christian Theiler, et~al.
\newblock Real-time feedback control of the impurity emission front in tokamak divertor plasmas.
\newblock {\em Nature communications}, 12(1):1105, 2021.

\bibitem{Timedep2}
Yoeri Poels, Gijs Derks, Egbert Westerhof, Koen Minartz, Sven Wiesen, and Vlado Menkovski.
\newblock Fast dynamic 1d simulation of divertor plasmas with neural pde surrogates.
\newblock {\em Nuclear Fusion}, 63(12):126012, 2023.

\bibitem{Timedep3}
JD~Lore, S~De~Pascuale, P~Laiu, B~Russo, J-S Park, JM~Park, SL~Brunton, JN~Kutz, and AA~Kaptanoglu.
\newblock Time-dependent solps-iter simulations of the tokamak plasma boundary for model predictive control using sindy.
\newblock {\em Nuclear Fusion}, 63(4):046015, 2023.

\bibitem{NormeyRico2007}
J.~E. Normey-Rico and E.~.F Camacho.
\newblock Control of dead-time processes.
\newblock {\em Advanced Textbooks in Control and Signal Processing London: Springer-Verlag}, 2007.

\bibitem{Seborg2016}
D.A.~Mellichamp D.E.~Seborg, T.F.~Edgar and F.J. Doyle.
\newblock Process dynamics and control, 4th~ed.
\newblock {\em John~Wiley \& Sons, Hoboken, NJ}, 2016.

\bibitem{Uytven2024}
Wim Van~Uytven, Wouter Dekeyser, Fabio Subba, Sven Wiesen, Niels Horsten, Nathan Vervloesem, and Martine Baelmans.
\newblock Discretization error estimation for eu-demo plasma-edge simulations using solps-iter with fluid neutrals.
\newblock {\em Contributions to Plasma Physics}, 64(7-8):e202300125, 2024.

\bibitem{Boeyaert2022}
Dieter Boeyaert, Stefano Carli, Kristel Ghoos, Wouter Dekeyser, Sven Wiesen, and Martine Baelmans.
\newblock Numerical error analysis of solps-iter simulations of east.
\newblock {\em Nuclear Fusion}, 63(1):016005, 2022.

\bibitem{thermalbifur}
Sergei~I Krasheninnikov and AS~Kukushkin.
\newblock Physics of ultimate detachment of a tokamak divertor plasma.
\newblock {\em Journal of Plasma Physics}, 83(5):155830501, 2017.

\end{thebibliography}


\begin{thebibliography}{9}

\bibitem{Zhu}
B Zhu et al., Phys. Plasmas 32 (2025) 062508

\bibitem{GuptaArXiv2025} 
A. Gupta, D. Eldon, E. Bang, K. Kwon, H. Lee, A. Leonard, J. Hwang, X. Xue, M. Zhao, and B. Zhu, 
“Machine learning surrogate model for detachment control in KSTAR,” arXiv:2505.07978 (2025).

\bibitem{Zhu2}
Zhu B, Zhao M, Bhatia H, et al. Data-driven model for divertor plasma detachment prediction. Journal of Plasma Physics. 2022;88(5):895880504. doi:10.1017/S002237782200085X

\bibitem{TCV}
T Ravensbergen et al., Nature Communications 12 (2021) 1105

\bibitem{DIII-D}
D Eldon et al., NF 57 (2017) 066039

\bibitem{KSTAR}
D Eldon et al., PPCF 64 (2022) 075002

\bibitem{EAST}
D Eldon et al., NME 27 (2021) 100963

\bibitem{MANTIS}
A Perek et al., Rev. Sci. Instrum. 90 (2019) 123514

\bibitem{Uytven2024}
Uytven, Discretization error estimation for EU-DEMO plasma-edge simulations using SOLPS-ITER with fluid neutrals, CPP 2024.

\bibitem{Boeyaert2022}
Boeyaert, Numerical error analysis of SOLPS-ITER simulations of EAST, NF 2022

\bibitem{thermalbifur}
Krasheninnikov SI, Kukushkin AS. Physics of ultimate detachment of a tokamak divertor plasma. Journal of Plasma Physics. 2017;83(5):155830501. doi:10.1017/S0022377817000654

\bibitem{cliff1}
Jaervinen et al., Phys. Rev. Lett. 121, 075001

\bibitem{cliff2}
A.G. McLean et al., Journal of Nuclear Materials 463 2015 533-536

\bibitem{cliff3}
H. Du et al., NF

\bibitem{cliff4}
X. Ma et al., NF

\bibitem{cliff5}
M. Zhao et al., NME 2024

\bibitem{Stangeby2018}
P C Stangeby 2018 Plasma Phys. Control. Fusion 60 044022

\bibitem{Body}
T. Body et al., Nuclear Materials and Energy 41 (2024) 101819

\bibitem{Goldston}
R J Goldston et al., PPCF 59 (2017) 055015

\bibitem{Eich2013NF}
T. Body et al., Nuclear Materials and Energy 41 (2024) 101819

\bibitem{XuPSI2025}
Xu et al., Nuclear Materials and Energy 41 (2024) 101819

\bibitem{XuNF2019}
Xu et al., NF 2019

\bibitem{ParkNF2018}
J Park et al., NF 2018

\bibitem{HFHD}
F Reimold et al., NME 2015

\bibitem{JET2015}
A Chankin et al., PPCF 57 095002 2015

\bibitem{JarvinenNME2017}
Jaervinen et al., "Interpretations of the impact of cross-field drifts on divertor flows in DIII-D with UEDGE" NME 12 1136 2017

\bibitem{MaNF2021}
X. Ma et al., "First evidence of dominant influence of E×B drifts on plasma cooling in an advanced slot divertor for tokamak power exhaust" NF 61 054002 2021

\bibitem{MaurizioNF2024}
R. Maurizio et al 2024 Nucl. Fusion 64 086048

\bibitem{ScottiNT}
F. Scotti et al. "Divertor characterization and access to dissipative divertor conditions in Negative Triangularity discharges in DIII-D" PPCF 2025

\bibitem{Timedep1}
Ravensbergen, T., van Berkel, M., Perek, A. et al. Real-time feedback control of the impurity emission front in tokamak divertor plasmas. Nat Commun 12, 1105 (2021).

\bibitem{Timedep2}
Yoeri Poels et al 2023 Nucl. Fusion 63 126012

\bibitem{Timedep3}
J.D. Lore et al 2023 Nucl. Fusion 63 046015

\bibitem{fopdt}
Seborg, Dale E., Thomas F. Edgar, and Duncan A. Mellichamp. Process Dynamics and Control. 4th ed. New York: Wiley, 2016.

\end{thebibliography}

\appendix

\section{Base model}
\label{appendix1}

For the choice of such base case, one needs to consider a balance between model accuracy and computational efficientcy, ensuring that the essential physics of divertor detachment are captured while maintaining fast and robust convergence. This balance is influenced by several key modeling setup:
\begin{itemize}
\item \textbf{mesh resolution} higher resolution leads to higher accuracy but slower convergence and can unnecessarily resolve small-scale dynamics not critical for detachment physics~\cite{Uytven2024,Boeyaert2022}.

\item \textbf{plasma transport models} the effects of cross-field drifts are important in forming detachment in the divertor, especially for medium-size tokamaks, however, it may increase the difficulty of converging a case and thus computational time, and reduce robustness.

\item \textbf{plasma fueling} 
    \begin{itemize}
        \item fueling by puffing deuterium gas through the outer boundary
        \item fueling by fixing deuterium ion density at the core boundary
    \end{itemize}

\item \textbf{impurity transport models}
    \begin{itemize}
        \item multi-charged model: Tracks the evolution of impurities across all charge states, providing realistic radiation dynamics. However, it requires solving additional impurity equations, increasing computational cost and sensitivity to initial conditions.
        \item fixed-fraction model: Assumes impurities remain in a prescribed fraction of the plasma, making it more robust and faster to converge, but may introduce artifacts.
    \end{itemize}
    
\end{itemize}


\subsection{Effects of mesh resolution in KSTAR detachment modeling}

\begin{figure}
  \centering
  \includegraphics[width=.9\textwidth]{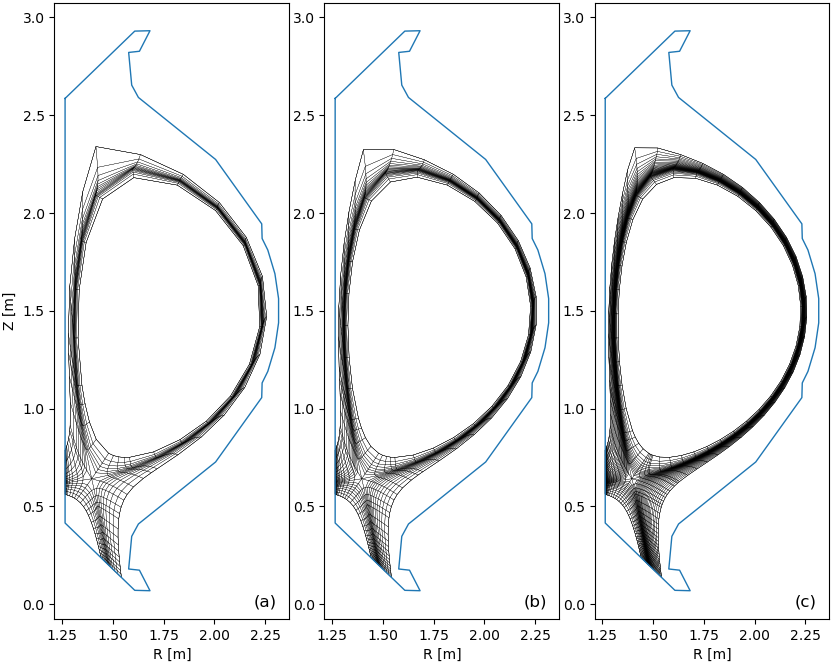}
  \caption{Three UEDGE mesh grids with different resolutions: the number of poloidal and radial cells at $48\times 14$ (a), $64\times 20$ (b), and $88\times 36$ (c).}\label{Fig:grid}
\end{figure}

\begin{figure}
  \centering
  \includegraphics[width=\textwidth]{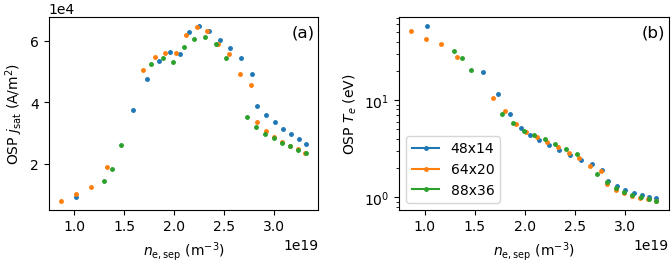}
  \caption{OSP ion saturation current $j_\mathrm{sat}$ (a) and OSP electron temperature $T_e$ versus outer midplane separatrix electron density $n_\mathrm{e,sep}$ for three different mesh resolutions: $48\times 14$ (blue), $64\times 20$ (orange), $88\times 36$ (green).}\label{Fig:reso}
\end{figure}

Three mesh resolutions, shown in Fig.~\ref{Fig:grid}, were tested to assess accuracy and computational efficiency. Density scans were carried out on each mesh with cross-field drifts included. All three cases produced similar results in terms of detachment onset and $j_\mathrm{sat}$ roll-over (see Fig.~\ref{Fig:reso}). The density scan with the lowest resolution required about $10-20$ minutes per converged case on average, while the medium-resolution mesh took roughly three times longer, and the highest resolution more than ten times longer. In order to maintain computational time under one hour, the medium-resolution mesh provides a good balance between accuracy and efficiency. Lower resolution suppresses oscillatory behavior and allows fewer steps to reach steady state. Therefore, the medium-resolution mesh (actually $64\times 24$) is chosen as the base case.

\subsection{Effects of drifts in KSTAR detachment modeling}

\begin{figure}
  \centering
  \includegraphics[width=\textwidth]{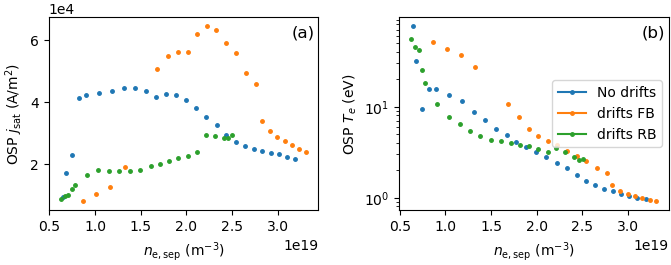}
  \caption{OSP ion saturation current $j_\mathrm{sat}$ (a) and OSP electron temperature $T_e$ versus outer midplane separatrix electron density $n_\mathrm{e,sep}$ for three different options for cross-field drifts: all drifts turned off (blue), all drifts turned on with ion $B\times\nabla B$ drift driven into the divertor (orange), and all drifts turned on with ion $B\times\nabla B$ drift driven out of the divertor (green).}\label{Fig:drift}
\end{figure}

Cross-field drifts play an important role in modeling detachment physics, particularly in capturing the in–out divertor asymmetry that arises from the $E\times B$ drift. However, including drifts may introduce numerical difficulties and increase computational cost. To evaluate whether it is worth including drifts in the database, we perform test cases with and without drifts and compare the results. As shown in Fig.~\ref{Fig:drift}, the inclusion of drifts has a significant impact on the results. In the forward $B_t$ configuration, detachment is more difficult to achieve, and the outer strike point $T_e$ is higher, primarily due to $E\times B$ transport of particles from the outer to the inner divertor. The added computational cost is modest, increasing the runtime by only about a factor of 2-3 on average. Therefore, it is justified to include drifts in the database.

\subsection{Effects of fueling methods in KSTAR detachment modeling}
\begin{figure}
  \centering
  \includegraphics[width=\textwidth]{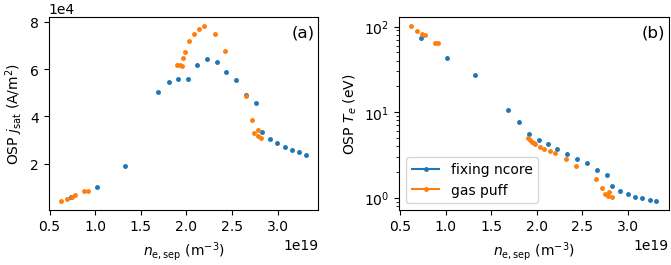}
  \caption{OSP ion saturation current $j_\mathrm{sat}$ (a) and OSP electron temperature $T_e$ versus outer midplane separatrix electron density $n_\mathrm{e,sep}$ for two methods of fueling: maintaining ion density at the core boundary (blue), puffing gas at the boundary of the outer midplane (orange).}\label{Fig:puff}
\end{figure}

In UEDGE, the plasma fueling can be implemented either through gas puffing or by fixing the ion density at the core boundary. A density scan comparing these two approaches is shown in Fig.~\ref{Fig:puff}, which indicates no significant differences in $j_\mathrm{sat}$ or $T_\mathrm{e,osp}$ for a given set of input parameters between the two methods. However, fixing the core boundary density proves to be numerically more stable and easier to converge. Moreover, the computational time required to achieve convergence with gas puffing is approximately three times longer than with fixed core boundary density. For these reasons, the database is generated using core-boundary-density fueling method.

\subsection{Effects of impurity models in detachment modeling}

In UEDGE, two models are available to handle impurity transport and radiation: the multi-charged model and the fixed-fraction model. The multi-charged model explicitly tracks all charge states of the impurity as well as the associated neutral gas, in a manner similar to SOLPS-ITER. In contrast, the fixed-fraction model provides a simplified treatment of impurity radiation. In this approach, radiation is calculated from pre-tabulated radiation rate coefficients as functions of the local electron density, temperature, and hydrogen neutral density for a given impurity species, assuming local charge balance. Importantly, impurity transport equations are not solved in this model, thereby excluding transport effects in order to reduce computational cost. Consequently, the impurity concentration profile must be specified as a user input.

\begin{figure}
  \centering
  \includegraphics[width=\textwidth]{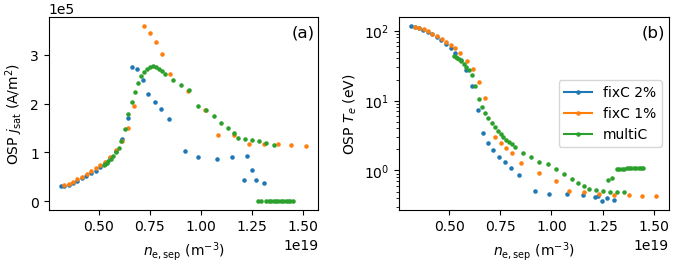}
  \caption{OSP ion saturation current $j_\mathrm{sat}$ (a) and OSP electron temperature $T_e$ versus outer midplane separatrix electron density $n_\mathrm{e,sep}$ using two impurity models: fixed fraction model of carbon assuming $1\%$ (orange) and $2\%$ (blue) carbon in the system, and multi-charged model of carbon (green).}\label{Fig:multi}
\end{figure}

A comparison between the two impurity models using a DIII-D equilibrium is presented in Fig.~\ref{Fig:multi} for a density scan with carbon impurities. The results show that the fixed-fraction model yields $j_\mathrm{sat}$ and $T_\mathrm{e,osp}$ profiles that are in good agreement with those of the multi-charged model when a reasonable carbon fraction is chosen ($1\%$ in this case). However, simulations with the multi-charged model require more than $20$ times longer to converge, making it impractical for generating the large number of runs needed for database production. For this reason, the fixed-fraction model is adopted in this work as a balance between accuracy and computational efficiency. For simplicity, we adopt a spatially uniform carbon fraction, and this fraction is treated as a scanned parameter, over a wide range to cover impurity concentrations relevant to KSTAR experiments.




\section{Database generation}
\label{appendix2}

A workflow framework has been developed to efficiently parallelize the execution of a large number of UEDGE simulations on high-performance computing (HPC) systems. This framework manages individual simulation tasks assigned to each CPU core, including determining appropriate initial conditions, monitoring convergence status, saving essential output data, and automatically launching the next case in the queue once the current simulation has successfully converged.

UEDGE employs a Jacobian-Free Newton-Krylov (JFNK) method with preconditioning to solve the plasma transport equations. The computational time required to reach a steady-state solution is highly sensitive to the choice of initial conditions. When the initial profiles of plasma variables (e.g., density, temperature) are close to the final steady-state solution, the Newton method can rapidly converge—sometimes within a single large time step. In contrast, if the initial profiles deviate significantly from the steady state, the solver may go through transient dynamics that increase computational time, despite these transients being of limited physical interest in this work. Consequently, assigning appropriate initial conditions for each case, especially when scanning over a large parameter space, is critical for generating a comprehensive database within a practical computational time.

Three key challenges must be addressed when generating the UEDGE database:
\begin{itemize}
    \item Convergence sensitivity to initialv conditions:
When the converged solution lies far from the initial condition, UEDGE often struggles to achieve convergence efficiently. Large deviations can lead to slow or unstable convergence behavior, sometimes requiring manual adjustments or user intervention to guide the solver.
    \item Numerical difficulty with extreme control parameters:
Obtaining solutions for control parameters outside the typical operational range—such as very low density combined with high input power—is challenging. In these regimes, transport is dominated by flux-limited dynamics, which tends to be more numerically unstable.
    \item Bifurcated solutions and hysteresis:
In scenarios involving possible bifurcations—such as thermal bifurcation~\cite{thermalbifur} or electron temperature cliff~\cite{cliff1,cliff2}—there is a risk that the solver may converge to the undesired branch. For this database, convergence to the second (detached-initiated) branch is not ideal, as experimental detachment typically evolves from the attached state along the first branch.

a bit of more comments on the hysteresis and control aspect.

\end{itemize}
To address the above challenges, the following steps are ingrained in the workflow:
\begin{enumerate}
    \item Manually generate an initial converged state:
Start by producing a single converged solution using control parameters near the center of the desired parameter space. This serves as a reliable starting point for subsequent runs.
    \item Generate a sparse database:
Run a small number of cases with sparsely sampled control parameter values across the intended range. These runs are allowed to proceed for a large number of steps to maximize the chance of convergence.
    \item Recover missing points manually:
For cases in the sparse database that fail to converge—especially those near the edges of the parameter space—attempt to obtain converged solutions manually if possible.
    \item Launch the full database sweep:
For each point in the full database, select a nearby initial condition from a previously converged case with lower collisionality, e.g. with lower density, lower input power, lower fraction of impurity or lower diffusivity. Set a conservative runtime or limit the number of time steps to avoid excessive computation. This approach helps to limit computational resources spent on cases that are unlikely to converge, e.g. cases with control parameters far outside the normal operational range.
\end{enumerate}

\end{document}